\documentclass[11pt]{article}
\usepackage{amsmath}
\usepackage{epsfig}
\usepackage{amsfonts}

\numberwithin{equation}{section}

\newcommand{\px}{\varphi} 
\newcommand{\ex}{\epsilon} 
\newcommand{\Ox}{\Omega} 
\newcommand{\gx}{\gamma}

\newcommand{\rep}[1]{\mathbf{#1}}
\newcommand{\nn}{\nonumber}
\newcommand{\cK}{\mathcal{K}}
\newcommand{\cF}{\mathcal{F}}
\newcommand{\cs}{\zeta}

\newcommand{\om}{\ensuremath{\omega}}

\newcommand{\N}{\ensuremath{{\cal N}}}

\newcommand{\ra}{\ensuremath{\rightarrow}}

\newcommand{\half}{\ensuremath{\frac{1}{2}}}

\newcommand{\quarter}{\ensuremath{\frac{1}{4}}}

\def\Vol{{\cal V}}

\newcommand{\ring}[1]{\mathring{#1}}
\def\w{\wedge}

\newcommand{\be}{\begin{equation}}
\newcommand{\ee}{\end{equation}}

\newcommand{\ba}{\begin{eqnarray}}
\newcommand{\ea}{\end{eqnarray}}

\newcommand{\ns}{\normalsize}

\begin{document}


\begin{titlepage}

\title{
   \hfill{\ns hep-th/0602163\\}
   \vskip 2cm
   {\Large\bf M-theory on seven-dimensional manifolds with $SU(3)$ structure}
\\[0.5cm]}
   \setcounter{footnote}{0}
\author{
{\ns\large 
  \setcounter{footnote}{3}
  Andrei Micu$^1$\footnote{email: amicu@th.physik.uni-bonn.de}~, 
  Eran Palti$^2$\footnote{email: e.palti@sussex.ac.uk}~, 
  P.M.Saffin$^{2,3}$\footnote{email: paul.saffin@nottingham.ac.uk}}
\\[0.5cm]
   $^1${\it\ns Physikalisches Institut der Universit\"at Bonn} \\
        {\it\ns Nussallee 12, D-53115, Bonn, Germany} \\[0.2em] 
   $^2${\it\ns Department of Physics and Astronomy, University of Sussex}\\
   {\it\ns Falmer, Brighton BN1 9QJ, UK} \\[0.2em] 
   $^3${\it\ns School of Physics and Astronomy, University of Nottingham}\\
   {\it\ns University Park, Nottingham NG7 2RD, UK}  
}
\date{}

\maketitle

\begin{abstract}\noindent
In this paper we study M-theory compactifications on seven-dimensional
manifolds with $SU(3)$ structure. As such manifolds naturally pick out a
specific direction, the resulting effective theory can be cast into a form
which is similar to type IIA compactifications to four dimensions. 
We derive the gravitino mass matrix in four dimensions and show
that for different internal manifolds (torsion classes) the vacuum preserves
either no supersymmetry, or $\N=2$ supersymmetry or, through spontaneous partial supersymmetry breaking, 
$\N=1$ supersymmetry. For the latter case we derive the effective $\N=1$ theory and give explicit 
examples where all the moduli are stabilised without the need of non-perturbative effects.

\vfill 
February 2006

\end{abstract}

\thispagestyle{empty}
\end{titlepage}

\section{Introduction}
\label{sec:introduction}

The low energy limit of M-theory, that is eleven-dimensional supergravity,
forms arguably the most natural starting point from which we hope to recover
observable physics from a fully consistent theory.  The first issue to address
is of course the fact that we observe four dimensions and the most
phenomenologically successful approach so far has been to single out one of
the space dimensions as independent of the other nine. Compactifying on this
dimension then leads to type IIA string theory
\cite{Campbell:1984zc,Huq:1983im,Giani:1984wc} which can then be compactified
to four dimensions on a six-dimensional Calabi-Yau.  The dimension may also be
taken to be an interval, and then compactifying on a Calabi-Yau leads to a
Brane-world scenario \cite{Lukas:1998yy}. If we do not require the existence
of such a special trivially fibred direction we should consider compactifying
on seven dimensional manifolds. The possible contenders for such manifolds are
required by supersymmetry to have special holonomy and until recently the main
body of work has concentrated on manifolds with $G_2$-holonomy that lead to
Minkowski space in four dimensions and preserve $\N=1$ supersymmetry
\cite{Papadopoulos:1995da}. These compactifications lead to massless scalar
fields in four dimensions that are known as moduli and an important first
phenomenological step is to lift these flat directions.  In string theory flux
compactifications have proved very successful in achieving this (for a review
see \cite{Grana:2005jc}) and in M-theory there has been some success in the
case of $G_2$-manifolds \cite{Acharya:2002kv,deCarlos:2004ci,Lambert:2005sh}.
A feature of flux compactifications is that flux on the internal manifold will
back-react on the geometry and in general induce torsion and warping on the
manifold deforming its special holonomy to the more general property of a G
structure \cite{Gauntlett:2002sc,Gauntlett:2003cy}.  To take this
back-reaction into account we should therefore consider compactifications on
manifolds with a particular G structure.  Compactifications that derive the
four dimensional theory have been done for the case of manifolds with $G_2$
structure \cite{Lambert:2005sh,House:2004pm,Dall'Agata:2005fm,D'Auria:2005rv}.
Eleven dimensional solutions that explore the structure of the vacuum have
been studied for the cases of $SU(2)$, $SU(3)$ and $G_2$ structure in
\cite{Kaste:2003zd,Dall'Agata:2003ir,Behrndt:2004mx,Lukas:2004ip,Behrndt:2004bh,Gauntlett:2004hs,Franzen:2005ve,Behrndt:2005im}.
An interesting point to come out of these studies is that compactifications on
manifolds with $SU(3)$ structure have a much richer vacuum spectrum than
manifolds with $G_2$ structure. Indeed there are solutions that preserve only
$\N=1$ supersymmetry in the vacuum putting them on an equal phenomenological
grounding with $G_2$ compactifications in that respect. There are however many
phenomenologically appealing features that are not present in the $G_2$
compactifications such as warped anti-deSitter solutions and solutions with
non-vanishing internal flux. 

In this paper we will study compactifications on manifolds with $SU(3)$
structure. We will see that because the $SU(3)$ structure naturally picks out
a vector on the internal manifold these compactifications can be cast into a
form that is similar to type IIA compactifications on $SU(3)$ structure manifolds \cite{House:2005yc}. 
However unlike in (massless) type IIA, we will show that it is
possible to find purely perturbative vacua with all the moduli stabilised that
preserve either $\N=2$ or $\N=1$ supersymmetry \cite{Behrndt:2004km,Behrndt:2004mj,Lust:2004ig}. 
Moreover, as also remarked in \cite{AFT}, such compactifications offer the possibility to obtain charged
scalar fields which reside in the $\N=2$ vector multiplets rather than in the
hypermultiplets as realised so far in most cases (see for example
\cite{Grana:2005jc}).

We will begin this paper with a discussion of the notion of G structures and
the idea of mass hierarchies between various G structures. In section
\ref{sec:reduction} we will perform a reduction of eleven-dimensional
supergravity on a general manifold with $SU(3)$ structure deriving the kinetic
terms for the four-dimensional scalar fields and the four-dimensional
gravitini mass matrix.  The mass matrix will then be used to explore the
amount of supersymmetry preserved by various manifolds. We will begin by
looking at vacua that preserve $\N=2$ supersymmetry in section
\ref{sec:n=2susy}. We will first derive the most general $\N=2$ solution and
use it as a check on the mass matrix. We will then show how this solution can
be used to find explicit vacua of an example manifold.  In section
\ref{sec:n=1} we will move on to the more phenomenologically interesting
$\N=1$ vacua and will show that some manifolds will induce spontaneous partial
supersymmetry breaking that will lead to an $\N=1$ effective theory. We will
derive this theory and go through an explicit example of moduli stabilisation.
This will also serve as an interesting example of a mass gap between
G structures. Finally, in the Appendices, we present our conventions and
some technical details related to the calculations we perform in the
main text.

\textbf{Note added:} While this manuscript was prepared for publication
another paper appeared, \cite{AZ}, which has some overlap with the issues
discussed in this paper. Further to this we were informed of work in progress which also 
relates to the discussed issues \cite{CL}.

\section{$G$ structures} 
\label{sec:gstructures}

In this section we briefly discuss the notion of a G structure and the two
particular cases of $G_2$- and $SU(3)$ structure in seven-dimensions. For a
more thorough introduction to G structures we refer the reader to
\cite{Gauntlett:2002sc,Gauntlett:2003cy}. A manifold is said to have
$G$ structure if the structure group of the frame bundle reduces to the group
$G$. In practice this translates into the existence of a set of $G$-invariant
forms and spinors on such manifolds.

In general these forms are not covariantly constant with respect to the
Levi-Civita connection, which would imply that the holonomy group of the
manifold is reduced to $G$. The failure of the Levi-Civita connection to have
reduced holonomy $G$ is measured by the \emph{intrinsic torsion}. In turn, the
intrinsic torsion, and in particular its decomposition in $G$-representations,
is used to classify such manifolds with $G$ structure. In the following we will
give a couple of examples of $G$ structures defined on seven-dimensional
manifolds which we will use in this paper.

\subsection{$G_2$ structure in seven dimensions}
\label{sec:g2}

A seven-dimensional manifold with $G_2$ structure has a globally defined
$G_2$-invariant, real and nowhere-vanishing three-form $\px$ which can be
defined by a map to an explicit form in an orthonormal basis \cite{joyce}.
Alternatively, manifolds with $G_2$ structure feature a globally defined,
$G_2$-invariant, Majorana spinor $\ex$. Note that we shall work in a basis
where Majorana spinors are real. In terms of this spinor the $G_2$ form, $\px$
is defined as
\begin{equation}
  \label{pxdef}
  \px_{mnp} = i \ex^T \gx_{mnp} \ex \; ,
\end{equation}
with the spinor normalisation $\ex^T \ex = 1$.

Using the $G_2$ structure form $\px$ we can write
\begin{equation}
  \label{g2torsion}  
  \begin{aligned}
  d\px = & \ W_1 \star \px - \px \w W_2 + W_3 \; , \\
  d\left( \star \px \right) = & \ \frac43 \star \px \w W_2 + W_4 \; ,
  \end{aligned}
\end{equation}
where $W_1, \ldots , W_4$ are the four torsion classes. In terms of $G_2$
representations $W_1$ is a singlet, $W_2$ a vector, $W_3$ a $\rep{27}$ while
$W_4$ transforms under the adjoint representation, $\rep{14}$. For further
reference we note here that manifolds with only $W_1 \neq 0$ are called
weak-$G_2$ manifolds and they are the most general solutions of the
Freund-Rubin Ansatz \cite{FR,BJ}.

\subsection{$SU(3)$ structure in seven dimensions}
\label{sec:su3}

Manifolds with $SU(3)$ structure are more familiar in the context of six
dimensions. In particular, the most important representatives are the
Calabi--Yau manifolds for which the intrinsic torsion vanishes identically
(ie, as explained before they have $SU(3)$ holonomy).  One the other hand,
seven-dimensional manifolds with $SU(3)$ structure were less studied partly
due to the fact that for the case of no torsion where the holonomy group of
the manifold is $SU(3)$ the seven-dimensional manifold is just a direct
product of a Calabi--Yau manifold and a circle. Therefore studying M-theory on
such manifolds is equivalent to studying type IIA string theory on a
Calabi-Yau. Once some torsion classes are non-vanishing a non-trivial
fibration is generated thereby making such studies different to type IIA
compactifications.

An $SU(3)$ structure on a seven dimensional manifold implies the existence of
two globally defined, nowhere-vanishing Majorana spinors $\ex_1$ and $\ex_2$
which are independent in that they satisfy $\ex_1^T \ex_2 = 0$. In the
following we will find it more convenient to use two complex spinors $\xi_\pm$
\begin{equation}
  \label{xipm}
  \xi_{\pm} = \frac1{\sqrt 2} \big(\epsilon^1\pm i\epsilon^2 \big)\; .
\end{equation}
Similar to the case presented in the previous subsection, we construct the
$SU(3)$ invariant forms $\Ox$, $J$, $V$
\begin{eqnarray}
  \label{OJVdef}
  \Ox_{mnp} & = & - \xi^{\dagger}_+\gamma_{mnp}\xi_- \nn \; ,\\
  J_{mn} & = & i \xi^{\dagger}_+\gamma_{mn}\xi_+ = -i \xi^{\dagger}_-\gamma_{mn}\xi_- 
  \; ,\\ 
  V_m & = & - \xi^{\dagger}_+\gamma_m\xi_+ = \xi^{\dagger}_-\gamma_m\xi_- \nn \; .
\end{eqnarray}
Note that in comparison to six-dimensional $SU(3)$ structures, in seven
dimensions there also exists a globally defined vector field $V$. It is
important to bear in mind that in general this vector is not a Killing
direction and thus the manifold does not have the form of a direct product. 

One can now show that $\Ox$, $J$ and $V$ are all the possible independent
combinations which one can construct and any other non-vanishing quantities
can be expressed in terms of them. For example we have
\begin{equation}
  \label{JwV}
  \begin{aligned}
    \xi^{\dagger}_-\gamma_{mnp}\xi_+ = & \ \bar \Omega_{mnp} \; ,\\
    \xi^{\dagger}_+\gamma_{mnp}\xi_+=\xi^{\dagger}_-\gamma_{mnp}\xi_- = & \ i(J\wedge
  V)_{mnp} \; .
  \end{aligned}
\end{equation}

Furthermore, one can also show that the forms defined in \eqref{OJVdef} satisfy
the seven-dimensional $SU(3)$ structure relations
\begin{equation}
  \label{su3rel}
  \begin{aligned}
  & J\wedge J\wedge J = \frac{3i}{4} \Omega\wedge \bar\Omega \; ,\\
  & \Omega\wedge J = V\lrcorner J = V \lrcorner \Omega = 0 \; ,
  \end{aligned}
\end{equation}
where the contraction symbol $\lrcorner$ is defined in equation \eqref{innpr}.
Finally one can prove the following useful relations
\ba
\label{ojvrel} 
V \lrcorner V &=& 1 \nn \; , \\
J^m_{\;\;i}J^i_{\;\;n}&=&-\delta^m_n+V^m V_n \nn \; ,\\
J_m^{\;\;i}\Omega_{\pm inp}&=&\mp\Omega_{\mp mnp} \; , \\
\star \Omega^{\pm} &=& \pm \Omega^{\mp} \wedge V  \; , \nn \\
\star \left( J\wedge V \right) &=& \half J \wedge J \; , \nn
\ea
where we have split the complex three-form $\Omega$ in to its real and
imaginary parts 
\be
\Omega = \Omega^{+} + i\Omega^{-} \; .
\ee

Let us now see how to decompose the intrinsic torsion in $SU(3)$ modules. As
before they are most easily defined from the differentials of the forms $\Ox$,
$J$ and $V$. Generically we have \cite{Dall'Agata:2003ir,Behrndt:2005im}
\ba
dV &=& R J + \bar{W_1} \lrcorner \Omega + W_1 \lrcorner \bar{\Omega} + A_1 + V
\wedge V_1 \; ,\\
dJ &=& \frac{2i}{3}\left( c_1 \Omega - \bar{c_1} \bar{\Omega} \right) + J
\wedge V_2 + S_1 +
V\wedge \left[ \frac{1}{3} \left( c_2 + \bar{c_2}\right) J +
  \bar{W_2}\lrcorner \Omega + W_2 \lrcorner \bar{\Omega} + A_2 \right] \; ,\\
d\Omega &=& c_1 J \wedge J + J \wedge T + \Omega \wedge V_3 + V \wedge \left[
  c_2 \Omega - 2 J \wedge W_2 + S_2 \right] \; , \label{su3torsion} 
\ea 
where the representatives of the 15 torsion classes are denoted by $R$,
$c_{1,2}$, $V_{1,2,3}$, $W_{1,2}$, $A_{1,2}$, $T$ and $S_{1,2}$. It is easy to
read off the interpretation of the above torsion classes in terms of the
$SU(3)$ structure group. There are three singlet classes $R$ (real) and
$c_{1,2}$ (complex), five vectors $V_{1,2,3}$ (real) and $W_{1,2}$ (complex),
three 2-forms $A_{1,2}$ (real) and $T$ (complex) and two 3-forms $S_{1,2}$.

Before concluding this section we should make more precise the relation
between the $SU(3)$ and $G_2$ structures on a seven dimensional
manifold. Obviously, as $SU(3) \subset G_2$, an $SU(3)$ structure
automatically defines a $G_2$ structure on the manifold. In fact, an $SU(3)$
structure on a seven-dimensional manifold implies the existence of two
independent $G_2$ structures whose intersection is precisely the $SU(3)$
structure. Concretely, using the spinor $\ex_1$ and $\ex_2$ defined above we
can construct the two $G_2$ forms $\px^{\pm}$
\begin{equation}
  \label{phi+-}
  \begin{aligned}
    \left( \px^{+} \right)_{mnp} \equiv & ~ 2i\epsilon_1 \gamma_{mnp} 
    \epsilon_1 \; ,  \\   
    \left( \px^{-} \right)_{mnp} \equiv & ~ 2i\epsilon_2 \gamma_{mnp} 
    \epsilon_2 \; .
  \end{aligned}
\end{equation}
The relation to the $SU(3)$ structure is now given by
\begin{equation}
  \label{phiOJV}
  \px^{\pm} = \pm \Omega^{-} - J \w V \; .
\end{equation}
Throughout this paper it will sometimes be useful to use the $SU(3)$ forms and
sometimes the $G_2$ forms but we should keep in mind that the two formulations
are equivalent.

\subsection{Mass hierarchies}
\label{sec:massstruct}

When the torsion on the internal manifold vanishes the holonomy group
directly determines the amount of supersymmetry preserved in the vacuum.
This is not the case with G structures where the amount of supersymmetry
in the vacuum need not be related to the structure of the manifold. It
should nevertheless be kept in mind that the amount of supersymmetry of
the effective action is not unrelated to the structure group. In
particular, the existence of globally defined spinors on the internal
manifold allows us to define four-dimensional supercharges and therefore
constitute a sufficient condition for supersymmetry of the effective
action. Even though in general the situation can be more complicated we
will assume that such supercharges, which are related to the globally
defined spinors, are the only ones which survive in four dimensions and so
the amount of supersymmetry of the effective action is given directly in
terms of the structure group of the internal manifold.\footnote{%
  We thank Nikolaos Prezas for pointing this out. For a recent discussion
  of this we refer the reader to \cite{Dall'Agata:2006vd}.}  Consequently,
we will consider that M-theory compactifications on seven-dimensional
manifolds with $SU(3)$ structure lead to an $\N=2$ supergravity theory in
four dimensions\footnote{%
  Strictly speaking, as manifolds with $G_2$ structure are known to have
  in fact $SU(2)$ structure \cite{FKMS}, the effective action in four
  dimensions would be that of an $\N=4$ supergravity. However, as $SU(2)$
  structures in seven-dimensions are much less tractable than $SU(3)$
  ones, we shall consider that the additional spinors lead to massive
  particles and we shall ignore them right from the beginning. In fact we
  shall see in sections \ref{sec:n=2susy} and \ref{sec:n=1} that for some
  seven-dimensional coset manifolds the $SU(2)$ structure is not
  compatible with the symmetries of the coset. As the lower mass states
  are associated with modes on the coset which obey the coset symmetries
  it is clear that such cases create a hierarchy between the four globally
  defined spinors effectively leading to a manifold with less globally
  defined spinors.}  while the vacuum may preserve $\N=2$ or $\N=1$
supersymmetry or even break it completely depending on which torsion
classes (and fluxes) are turned on.  This may be understood from the fact
that when there are more than one internal spinors on the manifold they
may satisfy different differential relations according to what torsion
classes are present and so may correspond to different eigenvalues of the
Dirac operator.  Consider decomposing the eleven-dimensional gravitino in
terms of the globally defined spinors on the internal manifold. Than the
four-dimensional gravitini may have varying masses and there will appear
mass hierarchies throughout the four-dimensional low-energy field
spectrum.  If the mass scales are well separated we can consider that only
the lowest mass states are excited and so it is clear that in such a
vacuum only a fraction of the original amount of supersymmetry is
preserved.  We will present such an example in section
\ref{sec:coset2weakg2} where it will become clear that one of the two
gravitini will become massive in the vacuum and thus supersymmetry will be
spontaneously broken from $\N=2$ to $\N=1$.

\section{The reduction}
\label{sec:reduction}

The theory we will be considering is the low energy limit of M-theory that is
eleven-dimensional supergravity.  The bosonic action of the theory as well as
the relevant gravitino terms are given by \cite{JP}
\ba
S_{11}&=&\frac{1}{\kappa_{11}^2}\int \sqrt{-g_{11}}d^{11}X\left[\half{\cal \hat{R}}_{11} 
        - \half\bar\Psi_{M}\hat\Gamma^{MNP}\hat D_{N}\Psi_{P}
        -\quarter\frac{1}{4!}\hat{F}_{MNPQ}\hat{F}^{MNPQ} \right. \label{mtheoryaction} \\\nonumber
        &~& \hspace{2cm} + \half\frac{1}{(12)^4}\epsilon^{LMNPQRSTUVW}\hat{F}_{LMNP}\hat{F}_{QRST}\hat{C}_{UVW} \\\nonumber
        &~& \hspace{2cm} - \left. \frac{3}{4(12)^2}( \bar\Psi_{M}\hat\Gamma^{MNPQRS}\Psi_{N}
                           +12\bar\Psi^{P}\hat\Gamma^{QR}\Psi^{S})
                                                F_{PQRS}\right] \; . \\ \nonumber
\ea
The field spectrum of the theory contains the eleven-dimensional graviton
$\hat{g}_{MN}$, the three-form $\hat{C}_{MNP}$ and the gravitino,
$\hat{\Psi}_{P}$.  The indices run over eleven dimensions
$M,N,..=0,1,...,10$. For gamma matrix and epsilon tensor conventions see
the Appendix.  $\kappa_{11}$ denotes the eleven-dimensional Planck
constant which we shall set to unity henceforth thereby fixing our units.

In this section we will consider this theory on a space which is a direct
product $M_{11} = M_4 \times K_7$ with the metric Ansatz
\be
ds_{11}^2 = g_{\mu \nu}(x) dx^\mu dx^\nu + g_{mn}(x,y) dy^m dy^n \label{metric},
\ee
where $x$ denotes co-ordinates in four-dimensions and $y$ are the co-ordinates
on the internal compact manifold. The first thing to note is that this Ansatz
is not the most general Ansatz possible for a metric as we have not included
as possible dependence of the four-dimensional metric on the internal
co-ordinates that is usually referred to as a warp factor. There are many
compactifications that can consistently neglect such a warp factor because
either a warp factor is not induced by the flux or it can be perturbatively
ignored if the internal volume is large enough. Including such a warp factor
is a difficult proposition for an action compactification because it can, and
generally will, be a function of the four-dimensional moduli \footnote{This is
  not a problem when looking for solutions as they only probe the vacuum and
  are insensitive to moduli dynamics.}.  For now we will proceed with an
unwarped Ansatz bearing in mind that this is only consistent for certain
compactifications.

The four-dimensional effective theory will be an $\N=2$ gauged supergravity.
These type of theories have been studied extensively in the literature, see
\cite{deWit:1984pk,Andrianopoli:1996vr,Andrianopoli:1996cm,DAASV,SV,ASV} and
references within, and this work will be useful as a guide for the
compactification.  In the upcoming sections we will derive most of the
quantities necessary to specify this theory. The kinetic terms for the low
energy fields will be derived from the Ricci scalar and the kinetic term for
the three-form. The prepotentials can then be derived from the
four-dimensional gravitini mass matrix.

\subsection{The Ricci scalar}
\label{sec:riccireducation}

As is well known, the metric on the compactification manifold is not rigid and
its fluctuations can be written in terms of scalar fields in the effective
low-energy theory. Important constraints on the spectrum and kinetic terms for
these scalar fields come from the fact that they should form a
four-dimensional $\N=2$ supergravity. Compactifications of type II
supergravities from ten to four dimensions on Calabi-Yaus naturally lead to
such a supergravity. In this section we will show that it is possible to keep
an analogy with these compactifications for the case of M-theory on
$SU(3)$ structure manifolds that we are considering. A similar approach was
adopted in \cite{House:2005yc} and we will closely follow their results.

\subsubsection{The induced metric variations}
\label{sec:inducedvariations}

Having $SU(3)$ structure on a manifold is a stronger condition than having a
metric.  Infact the $SU(3)$ structure induces a metric on the manifold that we
can write in terms of the invariant forms as
\begin{equation}
  \begin{aligned}
    g_{ab} \equiv & \; |s|^{-\frac{1}{9}} s_{ab} \; \, \\
    s_{ab} \equiv & \; \frac{1}{16} \left[\frac{1}{4} \left(\Omega_{amn}
        \bar\Omega_{bpq} + \bar\Omega_{amn} \Omega_{bpq} \right) 
      + \frac{1}{3} V_a V_b J_{mn} J_{pq} \right] J_{rs} V_t \;\;
    \hat\epsilon^{mnpqrst} \; .
  \end{aligned}
\end{equation}
Clearly, as the metric is determined uniquely in terms of the structure forms,
all the metric fluctuations can be treated as fluctuations of the structure
forms. The converse however is not true as it is possible that different
structure forms give rise to the same or equivalent metrics. Therefore, when
expressing the metric variations in terms of changes in the structure forms
one has to take care not to include the spurious variations as well.

Varying the formula above we can write the metric deformations as
\ba
 \delta g_{ab} &=& \frac{1}{8} \delta \Omega_{(a}^{\;\;\;mn} \bar{\Omega}_{b)mn} + \frac{1}{8} \Omega_{(a}^{\;\;\;mn} 
\delta \bar{\Omega}_{b)mn} 
 + 2V_{(a}\delta V_{b)} + V_a V_b \left(J\lrcorner \delta J \right) + J_{(a}^{\;\;\;m} \delta J_{b)m} \nonumber \\
 &\;& \; + V^m V_{(a}J^{n}_{\;\;b)} \delta J_{mn} 
 - \frac{1}{3} \left( \frac{1}{4} \delta \Omega \lrcorner \bar{\Omega} + \frac{1}{4} \Omega \lrcorner \delta \bar{\Omega} 
+ J \lrcorner \delta J \right) g_{ab} \; . \label{su3metvar}
\ea
Note that this is very similar to normal Calabi--Yau compactifications where
the metric variations were expressed in terms of K\"ahler class and complex
structure deformations. Keeping the terminology we will refer to the scalar
fields associated with $\delta J$ and $\delta \Omega$ as K\"ahler moduli and
complex structure moduli respectively. Furthermore we will denote the scalar
associated to $\delta V$ as the dilaton in complete analogy to the type IIA
compactifications.

Before starting the derivation of the kinetic terms associated to the metric
deformations discussed above we mention that the metric variations can be
dealt with more easily in terms of the variations of either of the two $G_2$
structures which can be defined on seven-dimensional manifolds with $SU(3)$
structure \eqref{phiOJV}
\be
\label{g2metricvar}
  \delta g_{ab} = \frac12 {\px^\pm_{(a}}^{mn} \delta \px^{\pm}_{b)mn} 
- \frac13 \left( \px^{\pm} \lrcorner \delta \px^{\pm} \right) g_{ab} .
\ee
Therefore, for each of the $G_2$ structures the formula coincides with the
metric variations on a manifold with $G_2$ structure \cite{House:2004pm}.

\subsubsection{The Ricci scalar reduction}
\label{sec:ricci4d}

Let us now see explicitely how to derive the kinetic terms for the moduli
fields described above. As they are metric moduli, their kinetic terms should
appear from the compactification of the eleven-dimensional Ricci scalar. The
explicit calculation is presented in Appendix \ref{app:ricci} and here we will
only outline the main steps before stating the final result. We should also
mention that during this process we are mainly interested in the fate of the
scalar fields which appear as fluctuations of the metric on the internal
manifold and therefore we shall not discuss the vector field (graviphoton),
which also arises from the metric, as we expect that its kinetic term is
the standard one.

For now we do not decompose $\Omega$ and $J$ into
their four-dimensional scalar components but with the vector $V$ we write
\be
V(x,y) \equiv e^{\hat{\phi}(x)} z(y) \label{Vdef},
\ee
where $z$ is the single vector we have on the internal manifold from the
$SU(3)$ structure requirements. Note that it is still $V$ and not $z$
that features in the $SU(3)$ relations (\ref{su3rel}). The difference
between $V$ and $z$ can be understood as $V$ is the $SU(3)$ vector
which also encodes the possible deformations of the manifold, while
$z$ is only a basis vector in which we expand $V$. Therefore, the
factor $e^{\hat \phi}$ encodes information about the deformations
associated to the vector $V$. This is completely analogous to the
compactification of eleven-dimensional supergravity on a circle to
type IIA theory and in order to continue this analogy we shall call
the modulus in equation \eqref{Vdef} the dilaton.
Let us further define a quantity which in the case where the
compactification manifold becomes a direct product of a
six-dimensional manifold (with $SU(3)$ structure) and a circle, plays
the role of the volume of the six-dimensional space
\begin{equation}
  \label{defv6}
  \Vol_6 \equiv e^{-\hat{\phi}} \Vol \; ,
\end{equation}
where $\Vol$ is the volume of the full seven-dimensional space
\begin{equation}
  \label{defv7}
  \Vol \equiv \int \sqrt{g_7} = \frac16 \int J \wedge J \wedge J \wedge V \; .
\end{equation}
To see the use of this quantity, note that due to the first relation in
\eqref{su3rel}, a scaling of the three-form $\Omega$ automatically induces a
change in the volume. Thus, scalings of $\Omega$ would have the same effect as
appropriate scalings of $J$ and in order not to count the same degree of
freedom twice we shall define
\begin{equation}
  \label{eqn:omegaCS}
    e^{\half K_{cs}}\Omega^{cs} \equiv \frac{1}{\sqrt{8}}\Omega
    (\Vol_6)^{-\half} \; , \\ 
\end{equation}
where we have also introduced the K\"ahler potential for the complex structure
deformations, $K_{cs}$, extending the results of \cite{House:2005yc,Grana:2005ny,Hitchin}
\begin{equation}
  \label{defocs}  
  K_{cs} \equiv - \mathrm{ln} \left( || \Omega^{cs} || \Vol_6
  \right) =-\mathrm{ln}i<\Omega^{cs}|\bar\Omega^{cs}> \equiv \int \Omega^{cs}
  \wedge \bar \Omega^{cs} \wedge z \; .
\end{equation}
It is easy to check that rescalings of $\Omega$ precisely cancel the
corresponding variation of $\Vol_6$ on the RHS of equation
\eqref{eqn:omegaCS} and hence $\Omega^{cs}$ defined on the LHS stays
unchanged. In this way we have managed to decouple the volume modulus
from the form $\Omega$. The relation \eqref{eqn:omegaCS} deserves one
more explanation. The additional factor on the LHS, $\exp{\frac12
  K_{cs}}$ has been introduced in order to describe by $\Omega^{cs}$
the exact analogue of the Calabi--Yau holomorphic 3-form whose norm
precisely gives the K\"ahler potential of the complex structure
deformations. 

One more comment is in order here. As explained before, not all the variations
of the structure forms induce valid metric deformations. In particular the
definition of the 3-form $\Omega$ \eqref{OJVdef} allows for an arbitrary phase
which would subsequently drop out from the metric variations
\eqref{su3metvar}. In order to make sure that such variations are not
introduced as degree of freedom we should ``gauge'' these phase transformation
for $\Omega$. Given the K\"ahler potential \eqref{defocs} and the definition
\eqref{eqn:omegaCS} it is not hard to see that K\"ahler transformations, which
correspond to scalings of $\Omega^{cs}$ by some function which is holomorphic
in the complex structure moduli, precisely correspond to phase variations of
$\Omega$. Therefore, the covariant derivative for the ``gauged'' phase
transformations of $\Omega$ should precisely be the K\"ahler covariant
derivative
\begin{equation}
  \label{Kcd}
  D_{\mu} \Omega \equiv \partial_\mu \Omega + \frac12 \partial_\mu K_{cs}
  \Omega = \sqrt {8 \Vol_6} e^{\frac12 K_{cs}} (\partial_\mu \Omega^{cs} +
  \partial_\mu K_{cs} \Omega^{cs}) \equiv \sqrt {8 \Vol_6} e^{\frac12 K_{cs}}
  D_{\mu} \Omega^{cs} \; .
\end{equation}

Finally we note that we have to take into account the usual Weyl rescalings in
order to arrive to the four-dimensional Einstein frame
\begin{equation}
  \label{weylrescaling}    
  \begin{aligned}
    g_{\mu\nu} \ra & ~\Vol^{-1} g_{\mu\nu} \; ,\\
    g_{mn} \ra & ~e^{-\frac{2}{3}\hat{\phi}} g_{mn}  \; .
  \end{aligned}
\end{equation}

Following the above steps one can derive the (linearised) variation of the
Ricci scalar under the metric fluctuation \eqref{su3metvar}. The calculation
is presented in the appendix and here we recall the final result
\begin{eqnarray}
  \label{R11su3}
  \int \sqrt{-g_{11}} d^{11} X \; \half {\cal R}_{11} & = & \int \sqrt{-g_4}
  d^4x \Big[ \half {\cal R}_4  - \partial_{\mu} \phi \partial^{\mu} \phi +
  \half e^{2\phi} \Vol^{-1} \int \sqrt{g_7} {\cal R}_7 \\ 
  & &\!\! -\frac{1}{8} e^{-\hat{\phi}} e^{K_{cs}} \int \sqrt{g_7}\; d^7y \; 
  D_{\mu} \Omega^{cs} \lrcorner D^{\mu} \bar{\Omega}^{cs}
  -\quarter \Vol_6^{-1} 
  e^{-\hat{\phi}} \int \sqrt{g_7} \; d^7y \; \partial_{\mu}J \lrcorner
  \partial^{\mu}J \Big] \; , \nn 
\end{eqnarray}
where we have also defined the four-dimensional dilaton
\begin{equation}
  \label{eqn:4Ddilaton}
  \phi \equiv \hat{\phi} - \half \mathrm{ln} \Vol_6 \; .
\end{equation}
The important thing to notice on this result is that the metric fluctuations
have naturally split into the dilaton, the $J$ and $\Omega^{cs}$ variations
with separate kinetic terms. Moreover, due to the dependence of $\sqrt{g_7}$
on the dilaton, it can be seen that the all the dilaton factors drop out from
the kinetic terms of the K\"ahler and complex structure moduli. Therefore,
this result is very much like the one for usual type IIA compactifications on
Calabi--Yau manifolds with the notable difference that a potential for the
moduli appears due to the fact that manifolds with $SU(3)$ structure are in
general no longer Ricci flat.

\subsection{Four-dimensional field content and kinetic terms}
\label{sec:kineticterms}

In this section we will complete the kinetic terms for the low energy scalar
field spectrum by reducing the three-form field $\hat C_3$. These scalar
fields pair up with the geometrical moduli into $\N=2$ multiplets. We will
however ignore the presence of additional fields, like gauge fields, which are
expected to have similar kinetic terms to the gauge fields coming from type
IIA compactifications. 

\subsubsection{Reduction of the three-form}
\label{sec:formfield}

As we have seen in the previous subsection, the compactification of the
gravitational sector of M-theory on seven-dimensional manifolds with $SU(3)$
structure resembles very much the corresponding compactifications of type IIA
theory on Calabi--Yau manifolds. Therefore we will find it useful to continue
this analogy at the level of the matter fields and so we will first decompose
the 3-form $\hat C_3$ along the vector direction which is featured in the
seven-dimensional manifolds with $SU(3)$ structure under consideration.
Consequently we write 
\be
\label{CB}
\hat{C_3} = C_3 + B_2 \w z \; ,
\ee
where $C_3$ is assumed to have no component along $z$, ie $C_3 \lrcorner z =
0$. As expected, in the type IIA picture $C_3$ will correspond to the RR
3-form, while $B_2$ represents the NS-NS 2-form field.
Then compactifying the eleven-dimensional kinetic term, taking care to
perform the appropriate Weyl rescalings \eqref{weylrescaling}, we arrive at
\ba
\label{C11su3}
& & \hspace{-1cm}\int{\sqrt{-g_{11}} d^{11} X \; \left[ -\quarter \hat{F}
    \lrcorner \hat{F} \right]} \\ 
& & = \int{\sqrt{-g_4} d^4x \left[ -\frac{1}{4} e^{2\phi} e^{-\hat{\phi}}
    \int \sqrt{g_7}d^7y \partial_{\mu} C_3 \lrcorner \partial^{\mu} C_3 -
    \quarter \Vol_6^{-1} e^{-\hat{\phi}} \int \sqrt{g_7}d^7y
    \partial_{\mu} B_2 \lrcorner \partial^{\mu} B_2 \right]} \; . \nn 
\ea
One immediately notices that the kinetic term for fluctuations of the
$B_2$-field along the internal manifold is the same as the kinetic term for the
fluctuations of the fundamental form $J$. Therefore we see
that these fluctuations pair up into the complex field
\be
T \equiv B_2 - iJ \; .\label{bigt}
\ee

In order to analyse the four-dimensional effective action we have to specify
which are the modes we want to preserve in a Kaluza-Klein truncation. In
general one restricts to the lowest mass modes, but in the case at hand this
is a hard task partly due to the big uncertainties regarding the spectrum of
the Laplace operator on forms for arbitrary manifolds with $SU(3)$ structure.
The best thing we can do is to use our knowledge from other similar cases
where the structure of four-dimensional theory was derived
\cite{House:2005yc,Grana:2005ny,Gurrieri:2002wz,Tomasiello:2005bp,D'Auria:2004tr}, as well
as the close analogy to the type IIA compactifications and postulate the
existence of a set of forms in which to expand the fluctuations we have
discussed so far. For the moment these forms are quite arbitrary, but for
specific cases it should be possible to derive some of their most important
properties. In fact we shall see such examples in sections \ref{sec:n=2susy}
and \ref{sec:n=1} where explicit examples of manifolds with $SU(3)$ structure
will be discussed.  Therefore we consider a set of two-forms, $\omega_i$, with
dual four-forms, $\tilde \omega^i$ which satisfy
\begin{equation}
  \label{defot}
  \int{\om_i \w \tilde{\om}^j \w z } = \delta^{j}_{i}  \; .
\end{equation}
Furthermore we introduce three-forms $(\alpha_A, \beta^A)$ which obey
\begin{equation}
  \label{defab}
  \begin{aligned}
    \int{\alpha_{A} \w \beta^{B} \w z} = & ~ \delta^{B}_{A} \; ,\\ 
    \int{\alpha_{A} \w \alpha_{B} \w z} = &~ \int{\beta^{A} \w \beta^{B} \w z}
    = 0 \; .
  \end{aligned}
\end{equation}
Anticipating that we expand the structure variations in these forms we also
consider them to be compatible with the $SU(3)$ structure relations
\eqref{su3rel} and \eqref{ojvrel}
\begin{equation}
  \begin{aligned}
    \om_{i} \w \alpha_{A} = \om_{i} \w \beta^{A} = & ~0 \; ,\\
    z \lrcorner \om_{i} = 
    z \lrcorner \alpha_{A} = z \lrcorner \beta^{A} = &~ 0\; . 
  \end{aligned}
\end{equation}
These forms can in general depend on all seven internal coordinates and not be
closed. The index ranges are not necessarily topological but should
correspond to the number of generalised calibrated submanifolds in the
internal manifold
\cite{Grana:2005ny,Gurrieri:2002wz,Tomasiello:2005bp,D'Auria:2004tr}.


Given the forms defined above we should expand all the fluctuations and
interpret the coefficients as the four-dimensional degrees of
freedom. Consequently we write for the metric variations
\begin{equation}
  \label{OJexp}
  \begin{aligned}
    J(x,y) = &~ v^i(x) \omega_{i}(y) \; ,\\
    \Omega^{cs}(x,y) =&~ Z^{A}(x) \alpha_{A}(y) - F_{A}\big(Z(x)\big)
    \beta^{A}(y) \; ,
  \end{aligned}
\end{equation}
where we have already used the fact that the deformations of $\Omega$ span a
special-K\"ahler manifold and therefore can be written as above, where $F_A$
is a holomorphic function of the complex coordinates $Z^A$, which is also
homogeneous of degree one in $Z^A$. From the four-dimensional perspective
$v^i$ are real scalar fields which we will refer to as K\"ahler moduli. $Z^A$
on the other hand are not all independent and we shall consider as the true
degrees of freedom the quantities $z^a = Z^a/Z^0$, where the index $a$ runs
over the same values as the index $A$, except for the value $0$.
For the matter fields we take
\begin{equation}
  \label{BCexp}
  \begin{aligned}
    B_2(x,y) = &~ \ring{B}_2(y) + \tilde{B}_2(x) + b^i(x) \omega_i(y) \; ,\\
    C_3(x,y) = &~ \ring{C}_3(y) + \tilde{C}_3(x) + A^i(x) \wedge \omega_i(y) + 
    \xi^{A}(x)\alpha_{A}(y) - \tilde{\xi}_{A}(x)\beta^{A}(y) \; .
  \end{aligned}
\end{equation}
Note that in the above decomposition we have allowed for a background value
for $B_2$ and $C_3$ which we denoted $\ring{B}_2$ and $\ring{C}_3$
respectively. These values should be understood as giving rise to the flux
terms for the field strengths of $B_2$ and $C_3$ and therefore they should not
be globally well defined over the internal manifold. We will postpone their
discussion until the next section when we deal with background fluxes. Note
that $B_2$ can not be expanded along the $z$ direction as it already comes
from a three-form with one leg along $z$, while $C_3$ was assumed not to have
any component along $z$ cf equation \eqref{CB}. The fields $b^i$, $\xi^A$ and
$\tilde \xi_A$ are scalar fields in four dimensions and they will be important
for our following discussion. Moreover, $\tilde{B}_2(x)$ is a four-dimensional
two-form which, in the absence of fluxes, can be dualised into an axion $b(x)$.
Here however we will not perform this dualization as in the examples we present
in sections \ref{sec:coset} and \ref{sec:coset2} the $\tilde{B}$-field will be massive
in four dimensions and therefore we will keep it as a member of ``the
universal'' tensor-multiplet.  $\tilde{C}_3(x)$ is a three-form which carries no degree
of freedom in four dimensions and is dual only to some constant, but its
dualisation in four dimensions requires more care. As explained before, we
shall not deal with the vector fields $A^i$ here as their couplings are
expected to be similar to the type IIA compactifications. Also we shall
neglect other vector degrees of freedom which arise from the isometries of the
internal manifold and leave their proper treatment for another project.

We will also find it useful to introduce at this level one more notation. As
we are mostly interested in the scalar fields in the theory we will denote all
the fluctuations of $\hat C_3$ which give rise to scalar fields in four
dimensions by $\hat c_3$. Just from its definition we can see that this is a
three-form on the internal manifold. In terms of the expansions above it takes
the form
\begin{equation}
  \label{hatc3}
  \hat c_3(x,y) = b^i(x) (\omega_i \wedge z)(y) + \xi^A(x) \alpha_A(y) - \tilde \xi_A(x) \beta^A(y)
  \; .
\end{equation}

Finally, as we expect that the low energy effective action is a $N=2$ (gauged)
supergravity, the light fields should assemble into $N=2$
multiplets. This is briefly reviewed in table \ref{N=2multiplets}. 
\begin{table}[h]
\center
 \begin{tabular}{||l|c||} 
   \hline

 $g_{\mu\nu}, A^0$   & gravitational multiplet  \\ \hline
 $\xi^0, \widetilde{\xi}_0, \phi, \tilde{B}_2$  & universal tensor-multiplet \\ \hline
 $b^i, v^i, A^i$ & vector multiplets \\ \hline
 $\xi^a, \widetilde{\xi}_a, z^a$  & hypermultiplets \\ \hline

\end{tabular}
\caption{Table showing the ${\cal N}=2$ multiplets}
\label{N=2multiplets}
\end{table}   
As mentioned before, the internal parts of the two form $B$, and the
fundamental form $J$ combine themselves into a complex field
\begin{equation}
  T(x,y)\equiv B_2(x,y) -iJ(x,y) = t^i(x) \omega_i(y) \equiv (b^i(x) - i v^i(x)) \omega_i(y) \; ,
\end{equation}
which will become the scalar components of the $N=2$ vector
multiplets. The associated K\"ahler potential is again similar to the
one in type IIA theory
\begin{equation}
  \label{vkahlerpot}
  K_{t} = -\ln{\frac16 \int J\wedge J\wedge J\wedge z} = -\mathrm{ln} \Vol_6 
  \; .
\end{equation}
As we expect from the structure of $\N=2$ supergravity theories as well as
from the analogy to type IIA compactifications \cite{House:2005yc,Grana:2005ny}, the fields
$t^i$ span a special K\"ahler geometry with a cubic prepotential $\mathcal {F}
= -\frac16 \; \frac{\cK_{ijk} t^i t^j t^k}{t^0}$, where $\cK_{ijk}$ are the
analogue of the triple intersection numbers
\begin{equation}
  \cK_{ijk} = \int \omega_i \wedge \omega_j \wedge \omega_k \wedge z \; .
\end{equation}
The symplectic sections are given by $X^I= (t^0, t^i)$ and $\cF_I = \partial_I
\cF$ with $t^0=1$. Indeed, one can easily check, using the expansion
\eqref{OJexp} that the K\"ahler potential above derives from the general
$\N=2$ formula $K=-\ln{i \big(X^I \bar \cF_I - \bar X^I \cF_I \big)}$.

It is interesting to note that while in type IIA compactifications with
fluxes only charged hypermultiplets can appear, in the case of M-theory
compactified on seven-dimensional manifolds with $SU(3)$ structure one can
also obtain charged vector multiplets as also remarked in \cite{AFT}.
Indeed it is not hard to see that provided $\int d \omega_i \lrcorner
(\omega_j \wedge z) \equiv k_{ij}$ does not vanish, the kinetic term for the
three-form $\hat C_3$ in eleven dimensions generates a coupling of the
type $k_{ij} b^i A^j$ in the low energy effective action which precisely
uncovers the fact that the scalars in the vector multiplets become
charged.

\subsection{Flux and gravitino mass matrix}
\label{sec:flux}

So far we have only discussed the kinetic terms of the various fields
which appear in the low energy theory and we have seen that their
structure is very much like in type IIA compactifications. We will now
turn to study the effect of the non-trivial structure group and of
turning on fluxes. The only background fluxes which can be turned on
in M-theory compactifications and which are compatible with
four-dimensional Lorenz invariance can be written as
\begin{equation}
  \left[ \hat{F_4} \right]_{\mathrm{Background}}= f\eta_{4}+{\cal G} \; . \label{fluxde}
\end{equation}
Here $f$ is known as Freud-Rubin parameter where $\eta_4$ is the
four-dimensional volume form and ${\cal G}$ is the four-form background
flux which can locally be written as
\begin{equation}
  \label{F4flux}
  \mathcal{G} = d \ring{C}_3(y) \; ,
\end{equation}
where $\ring{C}_3(y)$ is the background part of the three-form field $\hat
C_3$ which was defined in equation \eqref{BCexp}. As observed in the
literature \cite{Lambert:2005sh,House:2004pm,BW}, the Freund-Rubin
flux is not the true constant parameter describing this degree of
freedom. Rather one has to consider the flux of the dual seven-form
field strength $\hat F_7$ 
\begin{equation}
  \label{F7}
  \hat F_7 = d \hat C_6 + \frac12 \hat C_3 \wedge \hat F_4 \; ,
\end{equation}
which should now be the true dual of the Freund-Rubin flux. As can
be seen the $\hat F_7$ flux also receives a contribution from the
ordinary $\hat F_4$ flux. Therefore, in general, the Freund-Rubin flux
parameter is given by
\begin{equation}
  \label{ftolambda} 
  f = \frac{1}{\Vol} \left(\lambda + \half \int  \hat{c}_3 \wedge {\cal G} + \half \int  \hat{c}_3 \wedge d\hat{c}_3 
  \right) \; ,
\end{equation}
where $\lambda$ is a constant which parameterizes the 7-form flux.

On top of these fluxes which can be turned on for the matter fields
one has to consider the torsion of the internal manifold with $SU(3)$
structure which is also known as ``metric flux''. The
effects of the torsion can be summarised as follows. We have already
seen that the compactification of the Ricci scalar contains a piece
due to the non-vanishing scalar curvature of the internal
manifold. This is entirely due to the torsion as manifolds with
$SU(3)$ holonomy are known to be Ricci flat. Moreover, a non-trivial
torsion is associated with non-vanishing exterior derivatives of the
structure forms. If we insist that we expand the fluctuations of these
structure forms as in equation \eqref{OJexp} it is clear that the
expansion forms cannot be closed. Therefore, the presence of torsion
forces us to perform the field expansions in forms which are no longer
closed. Such forms will induce in the field strength of the three-form
$\hat C_3$ terms which are purely internal and which are -- from this
point of view -- indistinguishable from the normal fluxes and so the flux in (\ref{fluxde}) 
is modified to be the full field strength expression
\begin{equation}
  \hat F_4 = f\eta_{4}+{\cal G} + d \hat c_3 \; ,
\end{equation}
where the derivative should be understood as the exterior derivative on
the seven-dimensional manifold. However such ``induced'' fluxes are not
constant, but they depend on the scalar fields which arise from $\hat
C_3$. It is also worth noting at this point that provided these scalar
fields are fixed at a non-vanishing value in the vacuum, these vacuum
expectation values will essentially look like fluxes for $\hat F_4$ in
that specific vacuum. We will use this fact later on when we discuss
moduli stabilization.

As mentioned before, the effect of the fluxes and torsion is to ``gauge''
the $\N=2$ supergravity theory and induce a potential for the scalar
fields. These effects can be best studied in the gravitino mass matrix to
which we now turn.  In an $\N=1$ supersymmetric theory, the gravitino mass
is given by the Kahler potential and superpotential, while in an $\N=2$
theory we have a mass matrix which is constructed out of the Killing
prepotentials (electric and magnetic) that encode information about the
gaugings in the hyper-multiplet sector. Moreover, the same gravitino mass
matrix appears in the supersymmetry transformations of the
four-dimensional gravitini and therefore its value in the vacuum gives
information about the amount of supersymmetry which is preserved in that
particular case. This can also be understood from the fact that unbroken
supersymmetry requires vanishing physical masses\footnote{In AdS space,
  the mass parameter which appears in the Lagrangian is not the true mass
  of a particle. Therefore we use the terminology \emph{physical mass} in
  order to distinguish the true mass from the parameter which appears in
  the Lagrangian.}  for the gravitino and so, non-zero eigenvalues of the
gravitino mass matrix in the vacuum imply partial or complete spontaneous
supersymmetry breaking. In the case of partial supersymmetry breaking of
an $\N=2$ theory, the superpotential and D-terms of the resulting $\N=1$
theory are completely determined by the $\N=2$ mass matrix.

In a compactification from a higher-dimensional theory there are several
ways to determine the gravitino mass matrix in the four-dimensional theory.
If we have explicit knowledge of the four-dimensional degrees of freedom
we can derive the complete bosonic action and from the potential and gaugings
derive the $\N=2$ Killing prepotentials. Alternatively one can directly
perform a computation in the fermionic sector and directly derive the
gravitino mass matrix or compactify the higher dimensional supersymmetry
transformations. The advantage of the last two methods is that one obtains a
generic formula for the mass matrix in terms of integrals over the internal
manifold without explicit knowledge of the four-dimensional fields. Once these
fields are identified in some expansion of the higher-dimensional fields one
can obtain an explicit formula for the mass matrix which should also be
identical to the one obtained from purely bosonic computations. 

In the following we choose to determine the gravitino mass matrix by directly
identifying all the possible contributions to the gravitino mass from eleven
dimensions. For this we will first have to identify the four-dimensional
gravitini. Recall from section \ref{sec:su3} that on a seven-dimensional
manifold with $SU(3)$ structure one can define two independent (Majorana)
spinors which we have denoted $\ex_{1,2}$. Then, we consider the Ansatz
\be
\label{4dgravitini}
\hat{\Psi}_{\mu} =  \Vol^{-\quarter} \left( \psi^{1}_{\mu} \otimes \epsilon_1
  + \psi^{2}_{\mu} \otimes \epsilon_2 \right) \; ,
\ee
where $\psi^{1,2}$ are the four-dimensional gravitini which are Majorana
spinors and the overall normalisation factor is chosen in order to reach
canonical kinetic terms in four-dimensions. It is more customary to work with
gravitini which are Weyl spinors in four dimensions and therefore we decompose
$\psi^{1,2}$ above as
\be
\psi_{\mu}^{\alpha} = \half \left( \psi_{+\mu}^{\alpha} + \psi_{-\mu}^{\alpha}
\right) \; ,
\ee
where $\alpha, \beta = 1,2$ and the chiral components of four-dimensional
gravitini satisfy 
\be
\gamma_5\psi^\alpha_{\pm\mu} = \pm\psi^\alpha_{\pm\mu} \; .
\ee
Then compactifying the eleven-dimensional gravitino terms in
(\ref{mtheoryaction}) and performing the appropriate Weyl rescalings
\eqref{weylrescaling} we arrive at the four-dimensional action 
\begin{equation}
  \label{4dfermact}
  \tilde S_{\psi_\mu} = \int_{{\cal M}_4} \sqrt{-g} 
  \left[-\bar \psi^{\alpha}_{+\mu} \gamma^{\mu \nu \rho} D_\nu 
    \psi^{\alpha}_{+\rho} + S_{\alpha\beta} \bar{\psi}^{\alpha}_{+\mu }
  \gamma^{\mu\nu}\psi^{\beta}_{-\nu} + \mathrm{c.c.}\right] \; .
\end{equation}
The main steps in deriving the mass matrix are presented in appendix
\ref{app:massmatrix} and for similar calculations we refer the reader to the existing
literature \cite{House:2004pm,House:2005yc,AZ} where similar calculations were
performed. Equation \eqref{S}, which is the final result for the gravitino mass
matrix $S_{\alpha \beta}$, can be written as
\begin{eqnarray}
  \label{massmatrix}
  S_{11} &=& \frac{ie^{\frac{7}{2}\hat{\phi}}}{8\Vol^{\frac{3}{2}}} \left\{
    \int_{{\cal M}_7} \left[ dU^+ \w U^+ + 2 {\cal G} \w U^+ \right] +
    2\lambda \right\} \; ,\nonumber \\ 
  S_{22} &=& \frac{ie^{\frac{7}{2}\hat{\phi}}}{8\Vol^{\frac{3}{2}}} \left\{
    \int_{{\cal M}_7} \left[ dU^- \w U^- + 2 {\cal G} \w U^- \right] +
    2\lambda \right\} \; ,  \\ 
  S_{12} &=& S_{21} = \frac{ie^{\frac{5}{2}\hat{\phi}}}{8\Vol^{\frac{3}{2}}}
    \int_{{\cal M}_7} \left[ 2i{\cal G}\w \Omega^+ + 2i d \hat c \w \Omega^+ -
    2 dJ \w \Omega^+ \w z \right ] \; . \nn
\end{eqnarray}
Here $\mathcal{G}$ denotes the internal part of the background flux which was defined in equation
\eqref{F4flux}, $\lambda$ is the constant to which the three-form $\tilde{C}_3$ is
dual in four dimensions and we have further introduced 
\begin{equation}
  \label{eqn:N2field}
  U^{\pm}  \equiv  \hat c_3 + ie^{-\hat{\phi}}\phi^{\pm} = \hat c_3 \pm
  ie^{-\hat{\phi}}\Omega^- - iJ \w z \; ,
\end{equation}
where $\hat c_3$ denotes the purely internal value of the three-form field $\hat
C_3$ which which was defined in equation \eqref{hatc3}.

The diagonal terms in the mass matrix correspond to the gravitino masses for
separate compactifications on the two $G2$ structures. This follows from
associating each of the four-dimensional gravitini with one of the two
internal spinors in the $G_2$ forms (\ref{phi+-}). We can also read off the
prepotentials, $P^x$ and $Q^x$ for the hypermultiplets and the K\"ahler
potential, $K$, for the vector multiplets of the $\N=2$ supergravity by
comparing the mass matrix with the general expression for an $\N=2$ gauged
supergravity \cite{DAASV,SV,ASV}
\begin{equation}
  \label{genmassmatrix}  
  S_{\alpha\beta} = \frac{i e^{\half K} }{2} \sigma^x_{\alpha\beta} 
  \big(P_A^x X^A - Q^{xA} F_A \big) \; ,
\end{equation}
where $P^x_A$ and $Q^{xA}$ are the electric respectively magnetic
prepotentials which depend on the hypermultiplets in the theory while $(X^A,
F_A)$ is a symplectic section which characterizes the special K\"ahler
geometry of the vector multiplet scalars. Note that we have used the general
formula for the $\N=2$ gauged supergravity mass matrix which appears when both
electric and magnetic gaugings are present. This is because we expect to have
both type of gaugings which is in general signaled by the presence of massive
tensor multiplets in the four-dimensional effective action. It is easy to
infer that such massive tensors appear if one takes into account that the
one-form $z$, used in the expansion \eqref{CB}, is not closed. Squaring the
field strength which comes from this expansion, $B_2$ will pick up a mass
proportional to $\int dz \wedge \star dz$.

Finally we note that in a generic vacuum the off diagonal components of the
mass matrix are non-vanishing and therefore the gravitini as defined in
equation \eqref{4dgravitini} are not mass eigenstates. The masses of the two
gravitini are then given by the eigenvalues of the mass matrix evaluated in
the vacuum. If these masses are equal and the two gravitini physically
massless then supersymmetry is preserved in the vacuum. However this is not
the case in general and then one encounters partial (when one gravitino is
physically massless) or total spontaneous supersymmetry breaking. We shall
come back to this issue in section \ref{sec:n=1}.

\section{Preserving N=2 supersymmetry}
\label{sec:n=2susy}

In this section we will consider the case where the internal manifold is one
that will preserve the full $\N=2$ supersymmetry in the vacuum. We will begin
by studying the constraints such a solution should satisfy 
in section \ref{sec:solution}, moving onto studying the form of the mass matrix
for this solution in section \ref{sec:n=2massless}. Finally in section
\ref{sec:coset} we will go through an explicit example of such a vacuum by
considering the coset $SO(5)/SO(3)_{A+B}$.

\subsection{N=2 solution}
\label{sec:solution}

In this section we will classify the most general manifolds with
$SU(3)$ structure that are solutions to M-theory that preserve $\N=2$
supersymmetry with 4D spacetime being Einstein and admitting two Killing
spinors. In order to study such solutions in full generality we allow for a
warped product metric
\begin{equation}
  \label{warpedmetric}
  ds^2_{11} = e^{2A(y)}g_{\mu \nu}(x) dx^\mu dx^\nu + g_{mn}(x,y) dy^m dy^n \; ,
\end{equation}
but will eventually show that the warp factor, $A(y)$, vanishes.  This
class of solutions has also been recently discussed in
\cite{Behrndt:2005im}. We look for solutions to the eleven-dimensional
Killing spinor equation
\begin{equation}
  \nabla_M \eta + \frac{1}{288} \left[ \Gamma_M^{\;\;NPQR} - 8 \delta_M^{[N}
  \Gamma^{PQR]} \right] \hat F_{NPQR} \;\eta = 0 \; .
\end{equation}
For the background field strength $\hat F_{MNPQ}$ above we will consider
the most 
general Ansatz compatible with four-dimensional Lorentz invariance. Therefore,
the only non-vanishing components of $\hat F$ are $\hat F_{mnpq}$
and $F_{\mu \nu \rho \sigma} = f \epsilon_{\mu \nu \rho \sigma}$.

Given that the internal manifold has $SU(3)$ structure we know there exist at
least two globally defined Majorana spinors and so we take a killing spinor
Ansatz
\begin{equation}
  \eta = \theta_1(x) \otimes \epsilon_1(y) + \theta_2(x) \otimes \epsilon_2(y)
  \; .
\end{equation}
Since we are looking for $N=2$ solution we treat $\theta_1$ and $\theta_2$ as
independent. This will lead to more stringent constraints than the $\N=1$
case, where they may be related, which will make finding the most general
solution straightforward. As we are looking for four-dimensional maximally
symmetric spaces, the Killing spinors $\theta_{1,2}$ satisfy
\begin{equation}
  \label{4dks}
  \nabla_{\mu} \theta_{i} = - \frac{i}{2} \Lambda^i_1 \gamma_{\mu} \gamma
  \theta_{i} + \half \Lambda^i_2 \gamma_{\mu} \theta_{i}
  \;\;\;\;\;\mathrm{(no\ sum \ over \ } i) \; ,
\end{equation}
where the index $i=1,2$ labels the two spinors. The integrability condition
reads
\begin{equation}
  \label{einsteinsol} 
  R_{\mu\nu} = -3\left[ \left( \Lambda^i_1 \right)^2 + \left( \Lambda_2^i
  \right)^2 \right]g_{\mu\nu} \; , \quad i=1,2 \; ,
\end{equation}
and so one immediately sees that not all $\Lambda^i_{1,2}$ are independent,
but have to satisfy
\begin{equation}
  \left( \Lambda^1_1 \right)^2 + \left( \Lambda_2^1 \right)^2 = 
  \left(\Lambda^2_2 \right)^2 + \left( \Lambda_2^2 \right)^2 \; .
\end{equation}
Now decomposing the Killing spinor equation into its external and internal
parts we arrive at the following equations
\begin{eqnarray}
  \label{internalderiv}
  \nabla_m \epsilon_{1,2} &=& \left( \frac{i}{12}e^{-4A}f\gamma_m
  \right)\epsilon_{1,2} \; ,\\  
  0 &=& \left( \gamma_m^{\;\;npqr}{\hat F}_{npqr} - 8 \gamma^{pqr}
    {\hat F}_{mpqr} \right)\epsilon_{1,2} \; , \\ 
  \left( \frac{i}{2} \Lambda^{1,2}_1 \right)\epsilon_{1,2} &=& 
  \left( \half e^{A} \gamma^n \partial_n A + \frac{i}{6}e^{3A}f
  \right)\epsilon_{1,2} \; , \\ 
  \left( \frac{1}{2} \Lambda^{1,2}_2 \right)\epsilon_{1,2} &=& \left(
  -\frac{1}{288} e^{A} \gamma^{npqr}{\hat F}_{npqr}\right)\epsilon_{1,2} \; .
\end{eqnarray}
In order to classify this solution from the point of view of the $SU(3)$
structure we have find the corresponding non-vanishing torsion classes by
computing the exterior derivatives of the structure forms. Using their
definition in terms of the spinors \eqref{OJVdef} and applying the results
above one finds

\begin{eqnarray}
  \label{nowarping}
  d V &=& \frac{1}{3} f J \; ,\nonumber \\ 
  d J &=& 0 \; ,\\
  d\Omega &=& -\frac{2i}{3} f \Omega \wedge V \; , \nonumber \\
  dA &=& 0 \; . \nn
\end{eqnarray}
The first thing to note is that the warp factor $A$ is constant in this
vacuum and therefore can be set to zero by a constant rescaling of the
metric.
The second thing to observe, comparing with equation \eqref{su3torsion}, is
that only the singlet classes $R$ and $c_2$ are non-vanishing. Moreover, they
are not independent, but proportional to each other as they can both be
expressed in terms of the Freund-Rubin parameter $f$.

From equations \eqref{internalderiv} we can also determine the
parameters $\Lambda^i_{1,2}$, which determine the value of the
cosmological constant, which are given by
\begin{equation}
  \label{fsol}
  \begin{aligned}
    \Lambda^1_1 = & ~ \Lambda^2_1 = \frac{f}{3} \; , \\
    \Lambda^1_2 = &~  \Lambda^2_2 = 0 \; .
  \end{aligned}
\end{equation}

The Killing spinor equations \eqref{internalderiv} also give constraints on
the internal flux that imply it should vanish. However an easier way to see
this is to consider the integral of the external part of the
eleven-dimensional Einstein equation which reads 
\begin{equation}
  \int{R_{(4)}} + \frac{4}{3}\int{f^2} + \frac{1}{72}\int{{\hat F}_{mnpq} 
    {\hat F}^{mnpq}} = 0 \; .
\end{equation}
We see that substituting (\ref{fsol}) we indeed recover ${\hat F}_{mnpq}=0$.

Finally we note that in terms of the two $G_2$ structures $\px^{\pm}$,
equations \eqref{nowarping} can be recast into a simple form
\begin{equation}
  d \px^{\pm} = \frac{2}{3}f \star \px^{\pm} \; ,
\end{equation}
which shows that both $G_2$ structures are in fact weak $G_2$.

\subsubsection{The mass of the gravitini}
\label{sec:n=2massless}

We can now use this solution to illustrate the discussion on the relation between
the gravitini masses and supersymmetry and to check our form of the mass
matrix. Inserting the solution just derived into the mass matrix we should
find that the masses of the two gravitini degenerate and that they are both
physically massless. Taking the solution \eqref{nowarping} from the previous
section the mass matrix \eqref{massmatrix} reads 
\begin{equation}
  \label{SN2}
  \begin{aligned}
    S_{12} =& ~ 0 \; ,\\
    S_{11} =& ~ S_{22} = \frac{-ife^{\frac{7}{2}\hat{\phi}}}{3\Vol^{\half}}
    \; ,
  \end{aligned}
\end{equation}
which indeed shows that the masses of the two gravitini are the same. 
To show that the two gravitini are physically massless we recall that in AdS
space the physical mass of the gravitino is given by
\begin{equation}
  \label{mphys32}
  m_{phys} = m_{3/2} - l \; ,
\end{equation}
where $m_{3/2}$ is the actual mass parameter which appears in the Lagrangian
(in our case $|S_{11}|$), while $l$ is the AdS inverse radius and is defined
as
\begin{equation}
  R = -12 l^2 \; ,
\end{equation}
with $R$ the corresponding Ricci scalar.

In order to obtain the AdS radius correctly normalised we recall that the mass
matrix \eqref{SN2} was obtained in the Einstein frame which differs from the
frame used in the previous section by the Weyl rescaling
\eqref{weylrescaling}. Inserting this into \eqref{einsteinsol} we obtain the
properly normalised AdS inverse radius 
\begin{equation}
  \label{lN2}
  l = \frac{fe^{\frac{7}{2}\hat{\phi}}}{3\Vol^{\half}} \; .
\end{equation}
Note that here, as well as in equation \eqref{SN2}, the fields $\hat \phi$ and
$\mathcal{V}$ should be replaced with their particular values which they have
for this solution. 
Equation \eqref{lN2}, together with \eqref{SN2}, shows that the physical mass
of the gravitini, \eqref{mphys32}, vanishes confirming our expectations that
the vacuum determined in the previous section does indeed preserve $\N=2$
supersymmetry.

\subsection{The coset $SO(5)/SO(3)_{A+B}$}
\label{sec:coset}

In order to see the above considerations at work we will now go through an
explicit example of a manifold that satisfies the $N=2$ solution discussed in
the previous sections. The manifold we will consider is the coset space
$SO(5)/SO(3)_{A+B}$. Cosets are particularly useful as examples of structure
manifolds because the spectrum of forms that respect the coset symmetries is
highly constrained. There are more details about cosets in general and about
this particular coset in the appendix, or, for further reference we refer the
reader to \cite{KS}. In this section we summarise the results and construct a
basis of forms with which we can perform the compactification.
 
We begin by finding the most general symmetric two-tensor that respects the
coset symmetries, this will be the metric on the coset and is given by

\be
g = \left( \begin{array}{ccccccc} a  & 0  & 0  & 0 & d  & 0  & 0 \\
    0  & a  & 0  & 0 & 0 & d  & 0 \\
    0  & 0 & a  & 0 & 0 & 0 & d \\
    0  & 0  & 0  & b & 0 & 0  & 0 \\
    d  & 0  & 0  & 0  & c & 0  & 0 \\
    0  & d  & 0  & 0 & 0  & c & 0\\
    0  & 0  & d  & 0 & 0 & 0 & c \\
    \end{array}  \right) \; ,
\ee
where all the parameters are real. The parameters of the metric are the
geometrical moduli and we see that we have four real moduli on this coset.
Note that there is a positivity domain $ac > d^2 $. Having
established the metric on the coset we can move on to find the structure
forms. The strategy here is to find the most general one, two and three forms
and then impose the $SU(3)$ structure relations on them. It is at this stage
that we really see what the $G$ structure of the coset is.  This analysis is
performed in the appendix and we find that the structure forms are given by
\ba
V &=& e^{\hat{\phi}} z \; , \nn \\
J &=& v \; \omega \; \label{so5struct} , \\
\Omega &=&  \cs_3 \alpha_0 + \cs_4 \alpha_1 + \cs_6 \beta^1 + \cs_7 \beta^0 \; , \nn
\ea 
where the relations between the $\cs$s and the metric moduli are given in the appendix.
The basis forms satisfy the differential relations 
\begin{eqnarray}
  \label{so5basis} 
  dz &=& - \omega \; , \nn \\ 
  d\omega &=& 0 \; , \nn \\ 
  d\alpha_0 &=& z \wedge \alpha_1 \; , \nn \\ 
  d\beta^0 &=& - z \wedge \beta^1 \; , \\ 
  d\alpha_1 &=& 2 z \wedge \beta^1 - 3 z \wedge \alpha_0 \; , \nn \\
  d\beta^1 &=& -2 z \wedge \alpha_1 + 3 z \wedge \beta^0 \; . \nn
\end{eqnarray}
The structure forms (\ref{so5struct}) show that indeed the
coset has exactly $SU(3)$ structure. In terms of the moduli classification we
have been using it has a dilaton, one K\"ahler modulus and one complex
structure modulus\footnote{As is expected form $\N=2$ supergravity the
  parameters $\cs_3$,$\cs_4$,$\cs_6$ and $\cs_7$ describe only two real degrees of
  freedom.}
thus making up the four degrees of freedom in the metric.
We also show in appendix \ref{sec:coset} that scalar functions are in general
not compatible with coset symmetries and therefore we conclude that for such
compactifications no warp factor can appear.

\subsubsection{Finding $\N=2$ minima}
\label{sec:coset2structurevacuum}

In this section we want to find out if the potential which arises from the
compactification on the coset above has a minimum where the geometric moduli
are stabilised. In particular we wish to look for minima that preserve $\N=2$
supersymmetry and correspond to the solution discussed in section
\ref{sec:solution}. As usual, in a bosonic background, the condition for
supersymmetry is the vanishing of the supersymmetry variations of the
fermions. This is precisely what we used in the previous section and thus
a supersymmetric solution should satisfy all the conditions derived there,
and in particular \eqref{nowarping}. It is easy to see that the forms
\eqref{so5struct} obey 
\ba
\label{cosetderivs}
dV &=& -\frac{e^{\hat{\phi}}}{v} J \; , \nn \\
dJ &=& 0, \\ 
d \Omega  &=& z \wedge \left[ 
\left( -3 \cs_4 \right) \alpha_0 + 
\left( \cs_3 - 2 \cs_6 \right) \alpha_1 +
\left( 2 \cs_4 - \cs_7 \right) \beta^1 +
\left( 3 \cs_6 \right) \beta^0 \right].  \nn
\ea
Therefore these forms will in general not satisfy the solution constraints
(\ref{nowarping}). Requiring them to match the solution gives a set of
equations for the moduli that will exactly determine the value of the moduli
in the vacuum. For the coset at hand these are easy to solve and the solution
is given by
\begin{eqnarray}
  \label{N2sol}
  e^{\hat \phi} & = & \frac{6^{\frac13}\sqrt{42}}{14}
  \lambda^{\frac{1}{6}} \; , \nn \\  
  v & = & \frac{6^{\frac{2}{3}}}{7} \lambda^{\frac{1}{3}} ,\; \\
  \cs_3 &=& -\cs_6 = -i\cs_4 = i\cs_7 = \frac{6}{49}\left(i-1\right) \sqrt{7\lambda}, \; \nn
\end{eqnarray}
where we have replaced the Freund-Rubin flux $f$ by true flux parameter
$\lambda$ from equation \eqref{ftolambda}.  Note that this solution fixes all
the geometric moduli which is an important result for M-theory
compactifications. It is important to stress however that $\cs$ are not
the true complex structure moduli, but are related to them by the
rescaling \eqref{eqn:omegaCS}. However, the complex structure moduli
defined in \eqref{OJexp}, which can be most easily read off in special
coordinates, do not depend on the rescalings of $\Omega$ and therefore, in
our case the value of the single modulus is given by
\be
z^1 = \frac{Z^1}{Z^0} = \frac{\cs_4}{\cs_3} = i
\ee

It can also be shown that the other scalar fields, which
come from the expansion of the 3-form $\hat{C}_3$, \eqref{BCexp}, in the forms
\eqref{so5basis} are also stabilised. A simple argument to support this
statement is that non-vanishing values of these scalars would lead to a
non-zero internal ${\hat F}_4$ flux at this vacuum solution due to the
non-trivial derivative 
algebra the basis forms satisfy, \eqref{so5basis}, which in turn is ruled out
by the supersymmetry conditions found in section \ref{sec:solution}. Hence,
these scalar fields are forced by supersymmetry to stay at zero vacuum
expectation value and therefore are fixed.

It is also worth observing one more thing regarding this solution. If we think
in terms of the type IIA quantities we see that the K\"ahler modulus $v$ and
the dilaton $e^{\hat \phi}$ are not independent and choosing to stay in the
supergravity approximation on type IIA side, ie take $v \gg 1$, drives
the theory to the strong coupling regime which explains why such solutions
can not be seen in the perturbative type IIA approach.

Finally we note that as the solution above is supersymmetric, the
four-dimensional space-time is AdS with the AdS curvature which scales with
$\lambda$ as
\begin{equation}
  l \sim \frac{1}{\lambda^{\frac16}} \; .
\end{equation}
Thus, in the large volume limit (ie $\lambda \gg 1$) the
four-dimensional space-time approaches flat space.

\section{Preserving $N=1$ supersymmetry}
\label{sec:n=1}

In this section we will analyse the case where we only preserve $\N=1$
supersymmetry in the vacuum. We will show that this occurs due to spontaneous
partial supersymmetry breaking, much like in massive type IIA \cite{House:2005yc}, and that it is possible to write an effective
$\N=1$ theory about this vacuum. We will derive the K\"ahler potential and
superpotential for this theory and go through an explicit example of a
manifold that leads to this phenomenon.

\subsection{Spontaneous partial supersymmetry breaking}
\label{sec:susybreaking}

In section \ref{sec:flux} we showed that for certain manifolds there is a mass
gap between the two gravitini in the vacuum and if this is the case then the
vacuum no longer preserves the full $\N=2$ supersymmetry but rather
spontaneously breaks to either $\N=1$ or $\N=0$ supersymmetry the former
corresponding to one physically massless gravitino and the latter to no
massless gravitini.  In this section we will consider the case where the
vacuum still preserves $\N=1$ supersymmetry.  With this a mass gap of the
scale of supersymmetry breaking, which is set by the vev of the scalars,
appears throughout the spectrum and so we can consider specifying an effective
$\N=1$ theory that is composed of the lower mass states. The superpotential
and the K\"ahler potential for this theory will then be given by the mass of
the physically massless gravitino as is usual for $\N=1$ theories.
Determining the superfield spectrum is a more complicated problem and an
important role is played by constraints on general partial supersymmetry
breaking.

Partial supersymmetry breaking has been considered in
\cite{Cecotti:1984rk,Cecotti:1985sf,Ferrara:1995xi,Andrianopoli:2001gm,Louis:2002vy}.
Following their discussions we briefly summarise how the matter sector of the
theory is affected by the breaking.  In the ${\cal N}=2$ theory the fields
were grouped into multiplets as described in Table \ref{N=2multiplets}. Once
supersymmetry is broken these multiplets should split up into ${\cal N}=1$
multiplets. The ${\cal N}=2$ gravitational multiplet will need to split into a
${\cal N}=1$ 'massless' gravitational multiplet and a massive
spin-$\frac{3}{2}$ multiplet \cite{Louis:2002vy}
\be
\left( g_{\mu\nu},\psi_1,\psi_2,A^0 \right) \ra \mathrm{massless} \; \left(
  g_{\mu\nu},\psi_1\right) \;\; + \;\; \mathrm{massive} \; \left(\psi_2, A^0,
  A^1, \chi \right)
\ee
Here $A^1$ is a vector field which has to come from one of the vector
multiplets and $\chi$ is a spin-$\half$ fermion which come from a
hypermultiplet. Moreover, one also needs one Goldstone fermion and two
Goldstone bosons to be eaten by the gravitino and the two vector fields
respectively which become massive, and these additional Goldstone fields also
come from the hypermultiplet sector. Additionally, depending on the details of
the theory there will be a certain number of vector and hypermultiplets which
also become massive in this process. Integrating out all the massive fields one
is left with an $\N=1$ supergravity theory coupled to vector and chiral
multiplets. The scalar fields in an $\N=1$ theory span a K\"ahler manifold
which has to be a subset of the $\N=2$ scalar manifold. With the scalar fields
of the $\N=2$ vector multiplets the situation is quite simple as they are
already complex coordinates on a (special) K\"ahler manifold. However, for the
hyper-scalars this is not the case, and it is in general non-trivial to find
the right combinations which will represent the correct complex
coordinates. For simple cases, as we will encounter in this paper, this can be
done and one can find explicitely the correct complex combinations which span
the $\N=1$ scalar K\"ahler manifold.

Before concluding this section we should also mention some subtle issues
related to the spontaneous $\N=2 \; \to \N=1$ breaking. It has been shown
\cite{Cecotti:1984rk,Cecotti:1985sf,Ferrara:1995xi,Louis:2002vy,Gunara:2003td}
that in Minkowski space spontaneous partial supersymmetry breaking can only
occur if the symplectic basis in the vector-multiplet sector is such that no
prepotential exists. However these results do not apply to the cases we
discuss in this paper for the following reasons. First of all, the no-go result
above has been obtained for purely electric gaugings of the $\N=2$
supergravity. Here we will see that we encounter magnetic gaugings as well and
going to purely electric gaugings requires to perform some electric-magnetic
duality which, in special cases, can take us to a symplectic basis where no
prepotential exists. The second argument is that we will encounter the
phenomenon of spontaneous partial supersymmetry breaking in AdS space and in
such a case it is not clear how to extend the no-go arguments of
\cite{Cecotti:1984rk}.\footnote{We thank Gianguido Dall'Agata for useful
  discussions on this subject.}

\subsection{The superfields and K\"ahler potential}
\label{sec:nocomplex}

Although the general pattern of partial supersymmetry breaking is constraining
it is not enough to determine the superfields in general.  The particular
difficulty, as explained before, lies in truncating the hypermultiplet
spectrum by finding the appropriate K\"ahler submanifold. However for the
special case where we have only the universal hypermultiplet this is possible.
We will therefore restrict our general analysis to such a situation
anticipating also the fact that the specific example we will study in section
\ref{sec:coset2} will be of this type. In order to find models with only one
hypermultiplet we will rely on the observation of \cite{Micu:2004tz}, that
six-dimensional manifolds with $SU(3)$ structure for which $\Omega^+$ is exact
feature no complex structure moduli and therefore the hypermultiplet sector
corresponding to compactifications on such manifolds consists only of the
universal hyper-multiplet. We therefore restrict ourselves to the case where
the torsion classes in (\ref{su3torsion}) are restricted to
\begin{equation}
  \label{nocomplex} 
  \begin{aligned}
    \mathrm{Re}(c_1) = &~ V_2 = S_1 = c_2 = W_2 = A_2 = 0 \; ,\\
    \mathrm{Im}(c_1) \neq& ~0 \; ,
  \end{aligned}
\end{equation}
and we see that under these conditions that the three form $\Omega^{+}$ is
indeed exact. 

We further have to determine the gravitino mass matrix for this situation.
Using (\ref{massmatrix}), (\ref{eqn:4Ddilaton}), (\ref{defv6}) we find that in
the particular case considered above, \eqref{nocomplex}, the gravitino mass
matrix becomes diagonal due to the fact that the internal flux $\mathcal {G}$
has to be closed due to the Bianchi identity
\ba
\label{eqn:newSmatrix}
S_{11}&=&\frac{i}{8}\frac{e^{2\phi}}{\sqrt{\Vol_6}}\int _{{\cal M}_7}
[dU^+\wedge U^+ + 2 \mathcal{G} \wedge U^+ + 2\lambda] \; , \nn \\ 
S_{22}&=&\frac{i}{8}\frac{e^{2\phi}}{\sqrt{\Vol_6}}\int _{{\cal M}_7}
[dU^-\wedge U^- + 2 \mathcal{G}\wedge U^- + 2\lambda] \; , \\
S_{12}&=&S_{21}=0 \; . \nn
\ea
The condition (\ref{nocomplex}) appears to be quite strong and we have
already come across an example where this is violated in section
\ref{sec:coset}. On the other hand we know from
ref.~\cite{Behrndt:2005im} that an $\N=1$ anti-deSitter vacuum, which
is required for all the moduli to be stabilised, necessarily means
that $J$ is not closed. Hence we always expect at least one of the
torsion classes in (\ref{nocomplex}) to be non-vanishing. Other than
this we must take the condition as a limitation of this paper.

Let us now see how we can identify the surviving degrees of freedom in a
spontaneously broken $\N=2$ theory which comes from a compactification on a
manifold which satisfies the requirements above.  First of all we know that in
order to have partial susy breaking we need at least two Peccei-Quinn
isometries of the quaternionic manifold to be gauged such that the
corresponding scalar fields become Goldstone bosons which are eaten by the
graviphoton and another vector field in the theory. In the model at hand,
where we only have one hypermultiplet, we have three such shift symmetries
which can be gauged. They correspond to the axion, the dual of the two-form in
four dimensions, and the two scalar fields which arise from the expansion of
the three-form $\hat{c}_3$ in the basis of three-forms $(\alpha_0, \beta^0)$. In
order to gauge one of these last two directions, or a combination thereof, we
need that the corresponding combination of the forms $\alpha_0$ and $\beta^0$
is exact. Without loss of generality we will assume that $\beta^0$ is exact.
Consistency with equations \eqref{defot} and \eqref{defab} implies then that
$\alpha_0$ is not closed. We therefore see that the scalar field which comes
from the expansion in the form $\beta^0$, which we denote $\tilde \xi_0$, is a
Goldstone boson and will be eaten by one (or a combination) of the vector
fields which come from the expansion of $C_3$. Then the other Goldstone boson
can only be given by the dual of the two form $\tilde{B}_2$. The way to see how this
direction becomes gauged is obscured by the fact that we are dealing with a
two-form rather then directly with a scalar field, but we can note that
provided $z$ is not closed, but its derivative is proportional to one of the
two forms $\omega_i$, there will appear in the compactified theory a
Green-Schwarz interaction, $\tilde{B}_2 \wedge dA$, which upon dualization precisely
leads to the desired gauging.\footnote{The issue of the dualization is further
  obstructed by the fact that $B$ will be massive. This, as explained at the
  end of section \ref{sec:flux}, is triggered by the non-closure of the one
  form $z$, which leads to mass term for the two-form field $\tilde{B}_2$ of the type
  $\int dz \wedge * dz$.}  Therefore we learn that the fields which survive
the truncation in the $\N=1$ theory are the dilaton and the second scalar
field from the expansion of $\hat{c}_3$ which we denote by $\xi^0$. The final thing
which we need to do is to identify the correct complex combination of these
two fields which defines the correct coordinate on the corresponding K\"ahler
submanifold. Knowing that the $\N=2$ gravitino mass matrix becomes the
superpotential in the $\N=1$ theory, which has to be holomorphic in the chiral
fields, we are essentially led to the unique possibility
\begin{equation}
  \label{superu}
  U^{0\pm} \equiv \xi^{0} \pm ie^{-\phi}\left( \frac{-4iZ^0}{F_0}
  \right)^{\half} \; ,
\end{equation}
where the sign $\pm$ is determined by which of the gravitini is massless and
we will drop the index unless required for clarity. $Z^0$ and $F_0$ are the
coefficients of the expansion of $\Omega$ in the basis $(\alpha_0, \beta^0)$,
\eqref{OJexp}, and the quantity $-4iZ^0/F_0$ is a positive real number as in
the particular choice of symplectic basis we have made ($\beta_0$ is exact)
$Z^0$ is purely imaginary.

To check that this is indeed the correct superfield we should make sure we recover the moduli space metric from the K\"ahler potential in 
the gravitino mass. The appropriate kinetic terms in (\ref{C11su3}) read 
\ba
S_{kin}^{U} =   \int \sqrt{-g} d^{4}x \left[ - \left( \frac{F_0}{-4iZ^0} \right)e^{2\phi} \partial_\mu
\left(\xi^0 + i e^{-\phi} \left( \frac{-4iZ^0}{F_0}\right)^\half \right)
  \partial^\mu \left( \xi^0 - i e^{-\phi} \left( \frac{-4iZ^0}{F_0}\right)^\half \right)  \right].
\ea
The gravitino mass in the $\N=1$ theory is given by the product of the
K\"ahler potential and the superpotential 
\be
M_{\frac{3}{2}} = e^{\half K} |W|.
\ee
From this we can use (\ref{eqn:newSmatrix}) to read off the K\"ahler potential
\be
\label{n=1kahler}
e^{K/2} = \frac{e^{2\phi}}{\sqrt{8\Vol_6}} ,
\ee
It is then easily shown that indeed the superfield and K\"ahler potential
satisfy  
\be
\partial_{U^0}\partial_{\bar{U^0}} \ln\left[ \frac{e^{4\phi}}{8\Vol_6} \right]
= -\left( \frac{F_0}{-4iZ^0} \right)e^{2\phi}.
\ee
Hence we have identified the correct superfield in the truncated spectrum. Determining the superfields arising from the $\N=2$ vector 
multiplets is a much easier task as they are just the natural pairing found in (\ref{bigt})
\be
t^i \equiv b^i - iv^i \label{supert},
\ee
where the index $i$ now runs over the lower mass fields.

\subsection{The superpotential}
\label{sec:superpotential}

The superpotential for the $\N=1$ theory can be read off from the gravitino
mass to be  
\be
W = \frac{i}{\sqrt{8}} \left\{ \int_{{\cal M}_7} \left[ dU^{\pm} \w U^{\pm}
  \right] + \mathcal {G} \wedge U^\pm + 2\lambda  \right\} \label{n=1super} ,
\ee
where again the $\pm$ sign is fixed by the lower mass state. From this
expression for the superpotential we can see that we should generically expect
a constant term $\lambda$, linear terms in $U$, quadratic terms $~t^2$, $~U^2$
as well as mixed terms $~tU$. These type of potentials will, in general,
stabilise all the moduli and we will see such an example in the next section.

It is instructive to note that finding a supersymmetric solution for this
superpotential automatically solves the equations which are required for a
solution of the full $\N=2$ theory to preserve some supersymmetry.
Therefore, for such a solution, it would be enough to show, using the mass
matrix \eqref{eqn:newSmatrix}, that a mass gap between the two gravitini forms
in order to prove that partial supersymmetry breaking does indeed occur.

\subsection{The Coset $SU(3) \times  U(1)/U(1) \times U(1)$}
\label{sec:coset2}

In this section we will go through an explicit example of a manifold that
preserves $\N=1$ supersymmetry in the vacuum.  The manifold we will be
considering is the coset $SU(3) \times U(1) / U(1) \times U(1)$ and for
simplicity we shall turn off the four-form flux $\mathcal {G} =0$.  Details of
the structure of the coset can be found in the appendix and in this section we
summarise the relevant parts.  The coset is specified by three integers
$p$,$q$, and $r$ that determine the embeddings of the $U(1) \times U(1)$ in
$SU(3) \times U(1)$, where the integers satisfy
\be
0 \le 3p \le q \; ,
\ee
with all other choices corresponding to different parameterisations of the
$SU(3)$.  As with the previous coset example we can use the coset symmetries
to derive the invariant $SU(3)$ structure forms and the metric.  The metric is
given by
\begin{equation}
  g = \left( \begin{array}{ccccccc} 
      a & 0 & 0  & 0 & 0 & 0 & 0 \\ 
      0 & a & 0  & 0 & 0 & 0 & 0 \\  
      0 & 0 & b  & 0 & 0 & 0 & 0 \\  
      0 & 0 & 0  & b & 0 & 0 & 0 \\  
      0 & 0 & 0  & 0 & c & 0 & 0 \\  
      0 & 0 & 0  & 0 & 0 & c & 0 \\  
      0 & 0 & 0  & 0 & 0 & 0 & d \\  
    \end{array}  \right)\; ,
\end{equation}
where the parameters $a,b,c,d$ are all real. We can write the invariant forms
as  
\ba
\label{coset2su3} 
V &=& \sqrt{d} z \; ,\nonumber \\
J &=& a \omega_1 + b \omega_2 + c \omega_3 \; ,\\
\Omega &=& \sqrt{abc}\left( i \alpha_0 - 4 \beta^0 \right). \nn 
\ea
This basis can be shown to satisfy the following differential relations
\begin{eqnarray}
  \label{coset2diffrel}
  dz & = & m^i \omega_i \; , \nn \\
  d \omega_i & = & e_i \beta^0 \; , \quad d \tilde\omega^i = 0 \; , \\
  d \alpha_0 & = & e_i \tilde \omega^i \; , \quad d \beta^0 =0  \; , \nn
\end{eqnarray}
where we have introduced two vectors $e_i=(2,2,2)$, and $m^i=(\alpha, -\beta,
\gamma), ~ i = 1,2,3$ which encode the information about the metric
fluxes. The quantities $\alpha, ~\beta$ and $\gamma$ are not independent, but
satisfy $\alpha -\beta + \gamma=0$ and in terms of the integers $p$ and $q$
have the expressions
\begin{eqnarray}
  \alpha &\equiv& \frac{q}{\sqrt{3p^2 + q^2}} \; , \nonumber \\
  \beta &\equiv& \frac{3p+q}{2\sqrt{3p^2 + q^2}}\; ,  \\
  \gamma &\equiv& \frac{3p-q}{2\sqrt{3p^2 + q^2}} \; .\nn
\end{eqnarray}
This ends our summary of the relevant features of the coset. We see that this
manifold indeed has the required torsion classes (\ref{nocomplex}) and, as
expected, has no complex structure moduli and three K\"ahler moduli.

\subsubsection{$\N=1$ minimum}
\label{sec:coset2fterm}

As explained in \cite{Castellani:1983tc}, M-theory compactifications on the coset manifold
presented above are expected to preserve $\N=1$ supersymmetry in the
vacuum. Therefore we can use the machinery developed at the beginning of this
section and derive the $\N=1$ theory in the vacuum. We will also turn off the
four-form flux $\mathcal {G}$ and so,
using equations (\ref{n=1kahler}) and (\ref{n=1super}) we find the
superpotential and K\"ahler potential to be
\ba
W &=& \frac{1}{\sqrt{8}} \left[ 4 U^0 \left( t^1 + t^2 + t^3 \right) + 2 \alpha t^2 t^3 - 2 \beta t^1 t^3 + 2 \gamma t^1 t^2 + 
2 \lambda \right] ,\\
K &=& - 4 \mathrm{ln} \left[ -i\left( U^0 - \bar{U}^0 \right) \right] - \mathrm{ln} \left[ -i \left( t^1 - \bar{t}^1 \right)
 \left( t^2 - \bar{t}^2 \right)\left( t^3 - \bar{t}^3 \right)\right] + \mathrm{const.}
\ea
where the superfields $t^i$ were defined in \eqref{supert} while for $U^0$ we
have
\begin{equation}
  \label{Udef}
  U^{0\pm} = \xi^0 \pm i e^{-\phi} \; ,
\end{equation}
as (\ref{coset2su3}) gives $-4iZ^0/F_0=1$.
We can look for supersymmetric vacua to this action by solving the F-term
equations. For convenience we restrict to the family of cosets with $p=0$
though the results can be reproduced for more general choices of embeddings.
We find the solution to the F-term equations
\begin{equation}
  \label{n=1sol}
  \frac{t^1}{2} = t^2 = t^3 = U^0 = -i \sqrt{\frac{\lambda}{3}} \; .
\end{equation}
%

At this point we can go back to check which of the gravitini is more massive.
Inserting the solution (\ref{n=1sol}) into the expression of the mass matrix
(\ref{eqn:newSmatrix}) we obtain
\be
S_{11} > S_{22} \label{eabove}\; , 
\ee
which means $\psi^{2}$ is the lighter gravitino and the one that should be
kept in the truncated theory. This gravitino is physically massless as expected.
This also fixes the $\pm$ sign ambiguity in the
superfield and superpotential so that we have $U^0 \equiv U^{0-}$. Finally we
note that as this solution is a supersymmetric solution of the
truncated $\N=1$ theory and that according to (\ref{eabove}) the
gravitino masses are not degenerate we indeed have encountered the phenomenon
of partial super symmetry breaking.

\subsubsection{The structure in the vacuum}
\label{sec:coset2weakg2}

It is informative to look at the form of the G structure of the coset in the
vacuum in terms of the $G_2$ structures.  The two $G_2$ forms (\ref{phiOJV})
satisfy the vacuum differential and algebraic relations
\ba
d \px^{\pm} &=& \sqrt{2} \left( \frac{\lambda}{3} \right)^{\frac{3}{4}} \left[ -8 \beta^0 \w z \pm 2 \om_1 \w \om_2 
 + (\pm 2 + 1 ) \om_2 \w \om_3 \pm 2 \om_1 \w \om_3 \right] , \nonumber \\
\frac{2}{3}f\star \px^{\pm} &=& \sqrt{2} \left( \frac{\lambda}{3} \right)^{\frac{3}{4}} \left[ \pm 8 \beta^0 \w z - 2 \om_1 \w \om_2 
 - \om_2 \w \om_3 - 2 \om_1 \w \om_3 \right].
\ea
It is clear to see that only $\px^-$ is weak-$G2$, and this is indeed the
$G2$ structure that features in the superpotential and is associated with the
lower mass gravitino. This shows an explicit mass gap appearing between the
two $G_2$ structures which is the same mass gap that corresponds to the
partial supersymmetry breaking which we have used to write an effective $\N=1$
theory. Hence we have shown an example of the idea of an effective G structure
where we could have arrived at this truncated $\N=1$ theory through a
$G2$ structure compactification even though the manifold actually has
$SU(3)$ structure.  Finally we should note that we could have used the
condition that the manifold should be weak-$G_2$ in the vacuum to solve for
the values of the moduli in the vacuum as we did in section
\ref{sec:coset2structurevacuum} instead of solving the F-term equations.

\section{Conclusions}
\label{sec:conclusion}

In this paper we studied compactifications of M-theory on manifolds with
$SU(3)$ structure. We showed that these compactifications can be cast into a
form much like type IIA compactifications on six-dimensional manifolds with
$SU(3)$ structure. The classical potential for the fields in four-dimensions
differed however from the IIA case and we have proved in two explicit
examples that one can find vacua which fix all the moduli without the need of
non-perturbative effects. 

We have also shown that depending on the different torsion classes which can
be turned on for such manifolds one can arrange to preserve either $\N=1$ or
$\N=2$ supersymmetry. We have also argued that in the case of the $\N=1$
solution one encounters the phenomenon of partial supersymmetry
breaking. This arises due to the fact that the two spinors which define the
$SU(3)$ structure satisfy different differential relations -- or in other
words, they are eigenfunctions of the Dirac operator corresponding to
different eigenvalues -- leading in this way to different masses for the
corresponding gravitini. In such a case we have seen that effectively one can
ignore from the beginning one of the spinors which make up the $SU(3)$
structure leading in this way to a $G_2$-like compactification.

There are many interesting direction than can be followed from this paper. It
would be interesting to consider manifolds that are more general then the
restriction (\ref{nocomplex}) and in particular the case where both the $c_1$
and $c_2$ torsion classes are non-vanishing should lead to a theory with a
vacuum that preserves $\N=1$ supersymmetry and has a stable vacuum where the
axions are stabilised at non-zero values. This would correspond to the
unwarped solution with non-vanishing exact internal flux found in
\cite{Behrndt:2005im}.

We have not touched on the subject of realistic particle content in this paper
one reason being that one can not possibly achieve a viable spectrum of
particles in M-theory compactifications by considering smooth manifolds as we
do in this paper.  However, in the effort to construct four-dimensional
theories which contain chiral matter and gauge fields from M-theory
compactifications (for recent developments see \cite{ABL}), considering
seven-dimensional manifolds with $SU(3)$ structure should be very interesting
because, as shown in this paper one can easily fix all the bulk moduli. This
could be supplemented by turning on torsion classes that would lead to
off-diagonal terms in the mass matrix that can be interpreted as D-terms in
the effective $\N=1$ theory thereby breaking supersymmetry spontaneously.

\vspace{1cm}
\noindent
{\large\bf Acknowledgments} \newline \newline We would like to thank Pablo
Camara, Joseph Conlon, Gianguido Dall'Agata, Thomas House, Josef Karthauser,
Nikolaos Prezas and Silvia Vaul\`a for useful discussions. The work of PMS and
EP was supported by PPARC. The work of AM was supported by the European Union
6th Framework Program MRTN-CT-2004-503369 ``Quest for Unification'' and
MRTN-CT-2004-005104 ``ForcesUniverse''.

\appendix

\section{Conventions}
\label{app:conventions}

In this appendix we outline the conventions used throughout this paper. 
The index ranges are
\ba
M,N,P,Q,R,S,T,U,V,W &=& 0 ,...,10 \; , \nn \\
a,b,m,n,p,q,r,s,t &=& 0,...,6 \; , \\
\mu,\nu,\rho &=& 0,...,3 \nn \\
i,j,k &=& 1,..., \mathrm{Number\;of\;two-forms\;in\;the\;basis} \; ,\nn \\
A,B &=& 1,...,\mathrm{Number\;of\;three-forms\;in\;the\;basis} \; ,\nn \\
\alpha,\beta &=& 1,2 \; . \nn
\ea
We worked with a mostly plus metric signature 
\be
\hat{\eta}_{11}=(-1,+1,+1,...) \; ,
\ee
where generally $\hat{}$ denotes eleven-dimensional quantities.
The $\hat{\epsilon}$ tensor density is defined as
\be
\hat\epsilon_{0123...}=+1 \; ,
\ee
and we define the inner product between forms as
\be
\label{innpr}
\left( \omega_p \lrcorner \nu_q \right)_{\mu_{p+1}...\mu_q} \equiv
\frac{1}{p!}  \left( \omega_p \right)^{\mu_1...\mu_p} \left( \nu_q
\right)_{\mu_1...\mu_p\mu_{p+1}...\mu_q} \; .
\ee
The eleven-dimensional spinor conventions are such that the charge conjugation
operator is given by $\hat\Gamma_0$ 
\be
\bar{\hat{\Psi}} = \hat{\Psi}^\dagger\hat\Gamma_0 \; .
\ee
We decompose the eleven-dimensional gamma matrices as 
\ba
\hat\Gamma_\mu& =& \gamma_\mu\otimes 1 \nn \; ,\\
\hat\Gamma_m &=& \gamma_5\otimes \gamma_m \; ,
\ea
with $\gamma_m$ imaginary and $\gamma_\mu$ real and
\begin{equation}
  \begin{aligned}
    -i\gamma_{0123} = & \; \gamma_5 \; , \\
    \gamma_{01...6} = & -i \; .
  \end{aligned}
\end{equation}
%

\section{Ricci scalar reduction}
\label{app:ricci}

In this appendix we reduce the eleven-dimensional Ricci scalar using the
metric Ansatz (\ref{metric}).  Before we begin the calculation we should
comment on the kind of variations we consider here. In general,
seven-dimensional manifolds with $SU(3)$ structure can have isometries that
produce gauge fields in the effective lower dimensional theory. For the moment
we are not interested in such metric variations and only treat the scalar
modes which appear from the fluctuations of the metric on the internal
manifold. Moreover we are only interested in the lightest modes in the
Kaluza--Klein tower.  Thus we consider a metric, including the fluctuations,
of the following form
\begin{eqnarray}
  \label{metfluct}
  ds_{11}^2 & = & {\bar g}_{\mu \nu}(x) dx^\mu dx^\nu + {\bar g}_{mn}(x,y) dy^m
  dy^n \\
  & = & {\bar g}_{\mu \nu}(x) dx^\mu dx^\nu + [\bar g_{mn}^0(y) + {\bar
  h}_{mn}(x,y)] dy^m dy^n \; .\nn
\end{eqnarray}
Direct computation of the 11d Ricci scalar gives
\begin{eqnarray}
 & & \int{\sqrt{-g_{11}} d^{11} X \; \half R_{11}} \nonumber \\
 &=& \int \sqrt{-g_{11}} d^{11} X \half \left[{\bar R}_4 + {\bar R}_7 - {\bar g}^{mn} {\bar \Box}_4 {\bar g}_{mn}
  +\left(\frac34 {\bar g}^{mp} {\bar g}^{nq} - \frac14 {\bar g}^{mn} {\bar  g}^{pq} \right) \left( \partial {\bar g}_{mn}
  \right) \left(\partial {\bar g}_{pq} \right) \right] \nn \\
  & = & \int \sqrt{-{\bar g}_4} d^4 x \int \sqrt{\bar g} d^7 y \half \left[ {\bar R}_4 + {\bar R}_7 
  - \frac14 \left({\bar g}^{mp} {\bar g}^{nq} - {\bar g}^{mn} {\bar g}^{pq} \right) \left( \partial {\bar g}_{mn} \right) 
  \left(\partial {\bar g}_{pq} \right) \right] \; ,\nonumber
\end{eqnarray}
where in the last equation we have performed a partial integration with
respect to the four-dimensional integral. At this point we want to replace the
metric variations with variations of the structure forms. Although eventually
we wish to parameterise the variations in terms of the $SU(3)$ structure forms
at this point it is easier to work with the $G2$ forms. Using equation
(\ref{g2metricvar}) we arrive at
\begin{equation}
  \label{R11}
  \int \sqrt{-g_{11}} d^{11} X \; R_{11} = \int \sqrt{-{\bar g}_4} d^4 x \int
  \sqrt{\bar g} \left[ {\bar R}_4 + {\bar R}_7 - \frac{1}{12} (\partial
  \bar \varphi)_{mnp} (\partial \bar \varphi)^{mnp} + \frac32 \;
  \frac{(\partial \bar{\Vol})^2}{\bar{\Vol}^2} \right] \; ,
\end{equation}
where to reach this we used the $G2$ identities
\ba
\px_{m}^{\;\;pq}\px^{m}_{\;\;ab} &=& \left( \star \px \right)^{pq}_{\;\;ab} +
2 \delta^{pq}_{ab}  \; ,\nn \\
9\left( \star \px \right)^{[pq}_{\;\;\;[ab}\delta^{m]}_{n]} &=& \left( \star \px \right)^{pqmt}\left( \star \px \right)_{abnt}
 + \px^{pqm}\px_{abn} - 6 \delta^{pqm}_{abn} \; ,\nn \\
\px \lrcorner \delta \px &=& 3 \Vol^{-1} \delta \Vol \; ,
\ea
and the fact that only the symmetric part of $\px_{m}^{\;\;pq} \delta
\px_{npq}$ contributes to the gauge independent metric variations.  Here
$\bar{\Vol}$ is the volume of the internal manifold as measured with the
metric $\bar g_{mn}$ which thus contains the metric fluctuations.  Note that
because we only consider the lowest KK states, ${\bar R}_4$ is independent of
the internal coordinates and thus its integration produces a factor of the
seven-dimensional volume $\bar{\Vol}$.  In order to put the four-dimensional
action in the standard form we further need to rescale the four dimensional
metric as
\be 
{\bar g}_{\mu \nu} = \frac1{\bar{\Vol}} g_{\mu \nu} \; .
\ee
Apart from normalising the Einstein-Hilbert term correctly this rescaling
will also produce a term which precisely cancels the last term of
\eqref{R11}. Thus the final form of the compactified
eleven-dimensional Ricci scalar takes the form
\begin{equation}
  \label{R11fin}
  \int \sqrt{-g_{11}} d^{11} X \; R_{11} = \int \sqrt{-g_4} d^4
  x \Big[ R_4 + \int
  \sqrt{\bar g} \big( {\bar R}_7 - \frac{1}{12} (\partial
  \bar \varphi)_{mnp} (\partial \bar \varphi)^{mnp} \big) \Big] \; .
\end{equation}
At this stage we can move back to using the $SU(3)$ forms using the translation equation (\ref{phiOJV}). We also move to the string 
frame by rescaling the internal metric
\begin{equation}
  \label{gintstring}
  \bar g_{mn} =  e^{-\frac23 \hat \phi} g_{mn} \; ,
\end{equation}
where the dilaton is defined as in equation (\ref{Vdef}).
Defining the $SU(3)$ structure forms with respect to the metric $g_{mn}$ the
decomposition \eqref{phiOJV} becomes
\begin{equation}
  \bar \px^\pm = e^{-\hat \phi} (\pm \Omega^- - J\wedge V) \; .
\end{equation}
Before identifying the correct degrees of freedom in four dimensions, as
discussed in section \ref{sec:kineticterms} we need to take out the K\"ahler
moduli dependence from $\Omega$ and we do this by defining a 'six-dimensional'
volume $\Vol_6$ and the true 'holomorphic' three-form $\Omega^{cs}$ as in
equations (\ref{defv6}) and (\ref{defocs}). With these definitions we have
\begin{equation}
  \label{delphi}
  \partial \bar \px^\pm = e^{-\phi} \big( \pm \left(\partial \phi\right) e^{\half K_{cs}}\Omega_{cs}^- \pm
  \partial \left( e^{\half K_{cs}}\Omega_{cs}^- \right)-
  \frac{1}{\sqrt{\Vol_6}} \partial J \wedge V)  \; ,
\end{equation}
where it can be easily checked that 
\begin{equation}
  \label{delO}
  \left(\partial \left(e^{\half K_{cs}}\Omega^-_{cs}\right) \right)_{mnp}
  \left(e^{\half K_{cs}}\Omega^-_{cs}\right)^{mnp} = 0 \; ,
\end{equation}
and so when we square the expression (\ref{delphi}) there is no mixing between
the various terms. Inserting (\ref{delphi}) into (\ref{R11fin}) we arrive at
the final expression (\ref{R11su3}).

\section{The gravitini mass matrix}
\label{app:massmatrix}

In this appendix we will derive the four-dimensional gravtini mass matrix
through dimensional reduction of the appropriate terms in the
eleven-dimensional action. We wish to work in terms of the $SU(3)$ structure
quantities as defined in section \ref{sec:su3} and so we begin by writing the
eleven-dimensional gravitino ansatz (\ref{4dgravitini}) in terms of the
four-dimensional chiral gravitini (\ref{4dgravitini}) and the complex internal
spinors (\ref{xipm})
\be
\hat{\Psi}_{\mu} =  \Vol^{-\quarter} \left[ \left( \psi^{1}_{+\mu} + \psi^{1}_{-\mu} \right) \otimes \left( \eta_+ + \eta_- \right)  
 -i \left( \psi^{2}_{+\mu} + \psi^{2}_{-\mu} \right) \otimes \left( \eta_+ - \eta_- \right) \right] \; . \label{su3ans}
\ee
We now go through each term in (\ref{mtheoryaction}) that will contribute to
the four-dimensional mass matrix.

\paragraph{The kinetic term}

We begin with the eleven-dimensional kinetic term which will produce a mass term in four dimensions for the particular index range choices
\be
{\cal L}_1 = -\half\bar\Psi_{\mu}\hat\Gamma^{\mu n \nu}\hat D_{n}\Psi_{\nu} \; . \label{torterm}
\ee
This term is only non-vanishing when the internal spinors are not covariantely constant and so will correspond to the potential induced 
by the torsion on the manifold. To calculate this more precisely we use the relation for the covariant derivative acting on the spinors
\be
D_m \eta_{\pm} = \quarter \kappa_{mnp}\gamma^{np}\eta_{\pm} \;, \label{spinorderiv}
\ee 
where $\kappa_{mnp}$ is the contorsion on the internal manifold which is
anti-symmetric in its last two indices. 
Inserting (\ref{su3ans}) into (\ref{torterm}) and using (\ref{spinorderiv}) to evaluate the derivative on the spinors as well 
as (\ref{OJVdef}) to replace the spinor bi-linears with the $SU(3)$ forms we arrive at 
\ba
{\cal L}_1 &=& -\frac{1}{2\Vol^{\half}} \left\{ \bar{\psi^{1}}_{+\mu} \gamma^{\mu\nu} \psi^{1}_{-\nu} \left[  \frac{i}{2}\kappa_{[mnp]} 
\left(J \wedge V \right)^{mnp} -\frac{i}{2} \kappa_{[mnp]} \Omega^{-mnp} \right] \right. \nonumber \\
& &\;\;\;\;\; + \; \bar{\psi^{2}}_{+\mu} \gamma^{\mu\nu} \psi^{2}_{-\nu} \left[ \frac{i}{2}\kappa_{[mnp]} 
\left(J \wedge V \right)^{mnp} + \frac{i}{2} \kappa_{[mnp]} \Omega^{-mnp} \right] \\
& &\;\;\;\;\; + \; \bar{\psi^{1}}_{+\mu} \gamma^{\mu\nu} \psi^{2}_{-\nu} \left[ -i \kappa_{m[np]}V^{[n}\delta^{p]m} 
-\frac{i}{2} \kappa_{[mnp]} \Omega^{+mnp} \right] \nonumber \\
& &\;\;\;\;\; +  \left. \bar{\psi^{2}}_{+\mu} \gamma^{\mu\nu} \psi^{1}_{-\nu} \left[ i \kappa_{m[np]}V^{[n}\delta^{p]m} 
-\frac{i}{2} \kappa_{[mnp]} \Omega^{+mnp} \right] \; + \mathrm{c.c.}
\;\right\} \; . \nn \label{l1tor} 
\ea
Now using the identity
\be
\bar{\psi^{2}}_{+\mu} \gamma^{\mu\nu} \psi^{1}_{-\nu}  = \bar{\psi^{1}}_{+\mu}
\gamma^{\mu\nu} \psi^{2}_{-\nu} \; ,
\ee
we can see that actually the first terms in the third and fourth lines cancel. This can be reasoned from the fact that the mass matrix 
should be symmetric. Using (\ref{spinorderiv}) we can operate on the spinor bi-linears (\ref{OJVdef}) and derive the 
following useful relations
\ba
\label{torel} 
\left( dV \right)_{mn} &=& 2 \kappa_{[mn]p}V^p \; ,\nonumber \\
\left( dJ \right)_{mnp} &=& 6 \kappa_{[mn}^{\;\;\;\;\;\;r} J_{r|p]}\; , \\
\left( d\Omega \right)_{mnpq} &=& 12 \kappa_{[mn}^{\;\;\;\;\;\;r}
\Omega_{r|pq]} \; . \nn 
\ea
With this (\ref{torel}) we can eliminate the contorsion from (\ref{l1tor}) in
favour of differential relations of the structure forms and we obtain
\ba
{\cal L}_1 &=& -\frac{1}{2\Vol^{\half}} \left\{ \bar{\psi^{1}}_{+\mu} \gamma^{\mu\nu} \psi^{1}_{-\nu} 
\left[ \frac{i}{4} \left(dV\right)_{mn}J^{mn} + \frac{i}{96} \left( d\Omega^- \right)_{mnpq} \left( \star \Omega^- \right)^{mnpq} 
 + \frac{i}{12} \left( dJ \right)_{mnp} \left( \Omega^+ \right)^{mnp} \right] \right. \nonumber \\
& &\;\;\;\;\; + \; \bar{\psi^{2}}_{+\mu} \gamma^{\mu\nu} \psi^{2}_{-\nu} \left[ \frac{i}{4} \left(dV\right)_{mn}J^{mn} + \frac{i}{96} \left( d\Omega^- \right)_{mnpq} \left( \star \Omega^- \right)^{mnpq} 
 - \frac{i}{12} \left( dJ \right)_{mnp} \left( \Omega^+ \right)^{mnp} \right] \nonumber \\
& &\;\;\;\;\; + \; \bar{\psi^{1}}_{+\mu} \gamma^{\mu\nu} \psi^{2}_{-\nu} \left[ 
- \frac{i}{12} \left( dJ \right)_{mnp} \left( \Omega^- \right)^{mnp}\right] \nonumber \\
& &\;\;\;\;\; +  \left. \bar{\psi^{2}}_{+\mu} \gamma^{\mu\nu} \psi^{1}_{-\nu}
  \left[ - \frac{i}{12} \left( dJ \right)_{mnp} \left( \Omega^- \right)^{mnp}
  \right] \;  + \mathrm{c.c.} \; \right\} \; . \label{l1diff}
\ea
This concludes the reduction of the kinetic term and we now move on to the flux terms.

\paragraph{The flux terms}

We begin be reducing the term
\be
{\cal L}_2 = -\frac{1}{16}\bar\Psi^{\mu}\hat\Gamma^{\rho \sigma } \Psi^{\nu} F_{\mu\rho\sigma\nu}\; . \label{flux1term}
\ee
This term arises from the purely external Freud-Rubin flux which we write as in (\ref{fluxde}) and (\ref{ftolambda}). 
Then substituting (\ref{su3ans}) into (\ref{flux1term}) and after some gamma matrix algebra we arrive at
\be
\label{l2diff}
{\cal L}_2 = \left[ i\bar{\psi^{1}}_{+\mu} \gamma^{\mu\nu} \psi^{1}_{-\nu} + 
i\bar{\psi^{2}}_{+\mu} \gamma^{\mu\nu} \psi^{2}_{-\nu} + \mathrm{c.c.} \right] \left[ \frac{1}{4\Vol^{\frac{3}{2}}} 
\left( \lambda + \half \int  \hat{c}_3 \wedge {\cal G} \right)  \right] \; .
\ee
The second flux term reads 
\be
{\cal L}_3 = -\frac{3}{4(12)^2} \bar\Psi_{\mu}\hat\Gamma^{\mu\nu
  lmnp}\Psi_{\nu}F_{lmnp} \; . 
\ee
This is the term from the purely internal flux. Again the reduction is simple
and gives  
\ba
{\cal L}_3 &=&\frac{1}{4(12)^2\Vol^{\half}}  \left\{ \bar{\psi^{1}}_{+\mu} \gamma^{\mu\nu} \psi^{1}_{-\nu} 
\left[ F^{lmnp} \left( J \wedge V - \Omega^- \right)^{rst} \hat{\epsilon}_{lmnprst} \right] \right. \nonumber \\
& &\;\;\;\;\;\;\;\;\;\; + \; \bar{\psi^{2}}_{+\mu} \gamma^{\mu\nu} \psi^{2}_{-\nu} \left[ F^{lmnp} \left( J \wedge V 
+ \Omega^- \right)^{rst} \hat{\epsilon}_{lmnprst} \right] \\
& &\;\;\;\;\;\;\;\;\;\; + \; \bar{\psi^{1}}_{+\mu} \gamma^{\mu\nu} \psi^{2}_{-\nu} \left[ - F^{lmnp} \left( \Omega^+ \right)^{rst} 
\hat{\epsilon}_{lmnprst}\right] \nonumber \\
& &\;\;\;\;\;\;\;\;\;\; +  \left. \bar{\psi^{2}}_{+\mu} \gamma^{\mu\nu} \psi^{1}_{-\nu} \left[ - F^{lmnp} \left( \Omega^+ \right)^{rst} 
\hat{\epsilon}_{lmnprst} \right] \;  + \mathrm{c.c.} \; \right\} \; . \nn
\label{l3diff}  
\ea
Finally we recall that the purely internal flux has a contribution from the
the background flux $\mathcal{G}$, and one which is due to the torsion of the
internal manifold $d \hat{c}_3$, which combine into
\be
F_{lmnp} = {\cal G}_{lmnp} + \left( d\hat{c}_3 \right)_{lmnp} \; .
\ee
After performing the Weyl rescalings \eqref{weylrescaling}, the contributions computed above, \eqref{l1diff}, \eqref{l2diff}, and
\eqref{l3diff} yield the following mass terms for the gravitino in four
dimensions 
\begin{equation}
  \tilde{S}_{\mathrm{mass}} = \int_{{\cal M}_{11}} \sqrt{-g} \left[ {\cal L}_1 +
    {\cal L}_2 + {\cal L}_3 \right] = \int_{{\cal M}_4} \sqrt{-g} \left[
    S_{\alpha\beta} \bar{\psi}^{\alpha}_{+\mu } \gamma^{\mu\nu}
    \psi^{\beta}_{-\nu} + \mathrm{c.c.} \right] \; ,  
\end{equation}
where
\ba
\label{S}
S_{11} &=& -\frac{ie^{\frac{3}{2}\hat{\phi}}}{8\Vol^{\frac{3}{2}}} 
 \left\{ \int_{{\cal M}_7} \left[ d\Omega^- \wedge \Omega^- + dV \wedge V \wedge J \wedge J 
 + 2dJ \wedge \Omega^- \wedge V  \right.\right. \nonumber \\
 & & \hspace{30mm} - 2 {\cal G} \wedge \left( \hat{c}_3 + i \left( \Omega^- - J \wedge V\right) \right)
 - d\hat{c}_3 \wedge \hat{c}_3 \nonumber \\ 
  & & \left. \hspace{30mm} - 2i d\hat{c}_3 \wedge \left( \Omega^- - J \wedge V\right) \right] 
- 2 \lambda \big\} \; , \nonumber \\
S_{22} &=& -\frac{ie^{\frac{3}{2}\hat{\phi}}}{8\Vol^{\frac{3}{2}}} 
 \left\{ \int_{{\cal M}_7} \left[ d\Omega^- \wedge \Omega^- + dV \wedge V \wedge J \wedge J 
 - 2dJ \wedge \Omega^- \wedge V  \right.\right. \nonumber \\
 & & \hspace{30mm} - 2 {\cal G} \wedge \left( \hat{c}_3 + i \left( - \Omega^- - J \wedge V\right) \right)
 - d\hat{c}_3 \wedge \hat{c}_3 \nonumber \\ 
  & & \left. \hspace{30mm} - 2i d\hat{c}_3 \wedge \left( -\Omega^- - J \wedge V\right) \right] 
- 2 \lambda \big\}  \; , \nonumber \\
S_{12} = S_{21} &=& -\frac{ie^{\frac{3}{2}\hat{\phi}}}{8\Vol^{\frac{3}{2}}} \int_{{\cal M}_7} \left[ 
 2dJ \wedge \Omega^+ \wedge V - 2i e^{\hat\phi}{\cal G} \wedge \Omega^+
  - 2i e^{\hat\phi}d\hat{c}_3\wedge \Omega^+  \right] \; .
\ea
This action can be written in the form (\ref{massmatrix}) using (\ref{eqn:N2field}).

\section{Coset manifolds}
\label{app:coset}

In this appendix we wish to briefly describe the procedure through which we can derive explicit information on the coset such as 
the metric, the G structure forms and the basis forms and their differential relations.

Consider a compact group $G$ with some subgroup $H$ then we can decompose the Lie algebra as $g =H \oplus K$. So the Lie manifold 
${\cal M}_g$ is a fibration of the Lie manifold ${\cal M}_H$ over the base ${\cal M}_K$. The base manifold ${\cal M}_K$ is the coset 
manifold $\frac{G}{H}$. We now follow the discussion in \cite{Mueller-Hoissen:1987cq} and construct a set of Lie valued 
one-forms from elements on the fibre 
$L_y$ at a point $y$ on the coset manifold, which we then expand in terms of the generators of the groups $H$ and $K$
\be
\Theta \equiv L_y^{-1} d L_y \equiv \sigma^a H_a + e^i K_i \; ,
\ee
where the indices run over the number of generators of the subgroup. The forms $e^i$ will form the basis forms on the coset manifold and 
 using 
\be
d\Theta = dL^{-1} \w dL  = -L^{-1} dL \w L^{-1} dL = -\Theta \w \Theta \; ,
\ee
gives that the basis forms satisfy the differential relations 
\begin{equation}
  \label{cosetdiff}
  \begin{aligned}
    d\sigma^a = & -\half f^{a}_{\;\;bc}\sigma^b \w \sigma^c -\half
    f^{a}_{\;\;ij}e^i \w e^j \; ,\\ 
    de^i =& -\half f^{i}_{\;\;jk}e^j \w e^k - f^{i}_{\;\;aj}\sigma^a \w e^j \;
    , 
  \end{aligned}
\end{equation}
where $f$ are the structure constants of the group $G$.
These expressions allow us to calculate the differential relations on the coset. The useful property of the coset is that requiring 
$G$-invariance
\be
g L_y = L_{y'} h \; ,
\ee 
where $g \in G$ and $h \in H$, we recover the transformation rules for a basis form on the coset
\be
e^i(y') K_i = e^i(y) h K_i h^{-1} \; ,
\ee
which means that requiring homogeneity of the basis forms general $n$-tensor on the coset
\be
g = g_{i_1...i_n}  e^{i_1} \otimes ... \otimes e^{i_n} \; ,
\ee
should satisfy the relation
\be
f^{j}_{a i_1} g_{j i_2 ... i_n} + ... + f^{j}_{a i_n} g_{i_1 ... j} = 0\; ,
\;\;\; \forall a \; ,\label{cosetsym}
\ee
This is the expression that restricts the possible forms that respect the
coset symmetries which we can use to solve for the most general one-, two-, or
three-forms on the coset and also the metric. Having quickly derived the
relevant expressions (\ref{cosetdiff}) and (\ref{cosetsym}) we can move on to
consider the particular examples used in this paper. One immediate conclusion
we can draw from equation (\ref{cosetsym}) is that scalar functions that
correspond to $n=0$ must vanish. This is the general result that cosets can
not support warping.

\subsection{$SO(5)/SO(3)_{A+B}$}

The group $SO(5)$ has two commuting $SO(3)$ subgroups. Hence there are a number of ways to mod out the $SO(3)$ and the index $A+B$ 
refers to the case where the subgroup $H$ is taken to be a linear combination of the two $SO(3)$s. Then by calculating the structure 
constants and imposing (\ref{cosetsym}) we find that the most general symmetric two tensor on the coset, 
which we interpret as the metric, must take the form
\ba
g & = & a (e^1 \otimes e^1 +  e^2 \otimes e^2 + e^3 \otimes e^3) + b e^4
\otimes e^4 + c (e^5 \otimes e^5 + e^6 \otimes e^6 + e^7 \otimes e^7) \nn \\ 
&& + 2 d(e^{(1} \otimes e^{5)} + e^{(2} \otimes e^{6)} + e^{(3} \otimes
e^{7)}) \; ,
\ea
where all the parameters are real. Similarly, the most general one-, two-, and
three-forms are  
\ba
\label{cosetgenforms}
\Psi_1 &=& \cs_1 e^4 \;, \nn \\
\Psi_2 &=& \cs_2 \left( e^{15} + e^{26} + e^{37} \right) \; , \\
\Psi_3 &=& \cs_3 e^{123} + \cs_4 \left( e^{127} - e^{136} + e^{235} \right) +
 \cs_5 \left( e^{145} + e^{246} + e^{347} \right) \nn \\
  && \hspace{-.27cm} + \cs_7 e^{567} + \cs_6 \left( e^{167} - e^{257} + e^{356} \right) \; , \nn
\ea
where all the parameters can be complex. The structure forms $V$, $J$ and $\Omega$ must fall within the restrictions 
of (\ref{cosetgenforms}) and they can be uniquely determined by imposing the algebraic $SU(3)$ structure relations on the forms 
in (\ref{su3rel}). This leads to equations relating the complex parameters to the real metric moduli,
if we identify $\Psi_1$ with $V$, $\Psi_2$ with $J$, $\Psi_3$ with $\Omega$.
\ba
\cs_1 &=& \sqrt{b} \label{metricmodulisolution} \; , \nn\\ 
\cs_2 &=& \left( ac - d^2 \right)^{\half} \; , \nn\\
\cs_3 &=& \frac{\cs_6}{a^2} \left( d + i\left( ac - d^2 \right)^{\half}
\right)^2  \; , \nn  \\
\cs_4 &=& \frac{\cs_6 a}{\left( d + i\left( ac - d^2 \right)^{\half} \right)}
\; ,  \\
\cs_5 &=& 0 \; , \nn \\ 
\cs_6 &=& \frac{2\left( ac - d^2 \right)^{\half} a \sqrt{c} }{a + ic}  \; ,
\nn \\ 
\cs_7 &=& \frac{\cs_6 c}{\left( d - i\left( ac - d^2 \right)^{\half} \right)}
\; . \nn
\ea
Equations (\ref{metricmodulisolution}) give the form of $V$, $J$ and $\Omega$
and we see that the natural basis of forms on the manifold is 
\ba
z &\equiv& e^4 \; , \nn\\
\omega &\equiv& \left( e^{15} + e^{26} + e^{37} \right) \; ,\\ 
\alpha_0 &\equiv& e^{123} \;\;\; \beta^0 \equiv e^{567} \; , \nn\\ 
\alpha_1 &\equiv& \left( e^{127} - e^{136} + e^{235} \right) \;\;\; \beta^1
\equiv \left( e^{167} - e^{257} + e^{356} \right)  \; , \nn
\ea 
in terms of which we can write the forms as given in equation (\ref{so5struct}). The differential relations on the coset basis forms 
can be calculated using (\ref{cosetdiff}) and are given by 
\ba
d\sigma^1 &=& - \sigma^{23} - e^{23} - e^{67} \; , \nn \\ 
d\sigma^2 &=& \sigma^{13} + e^{13} + e^{57} \; , \nn \\ 
d\sigma^3 &=& - \sigma^{12} - e^{12} - e^{56} \; , \nn \\ 
de^1 &=& -\sigma^2 e^3 + \sigma^3 e^2 + e^{45} \; , \nn\\ 
de^2 &=& \sigma^1 e^3 - \sigma^3 e^1 + e^{46} \; ,\\ 
de^3 &=& -\sigma^1 e^2 + \sigma^2 e^1 + e^{47} \; , \nn\\
de^4 &=& -e^{15} - e^{26} - e^{37} \; , \nn\\
de^5 &=& -\sigma^2 e^7 + \sigma^3 e^6 + e^{14} \; , \nn \\
de^6 &=& \sigma^1 e^7 - \sigma^3 e^5 + e^{24} \; , \nn \\ 
de^7 &=& -\sigma^1 e^6 + \sigma^2 e^5 + e^{34} \; , \nn 
\ea
From these expressions it is easy to calculate the basis form differential relations (\ref{so5basis}). 

\subsection{$SU(3)\times U(1)/U(1)\times U(1)$}

This coset was first studied in \cite{Castellani:1983tc}.
In this case we have $G=SU(3)\times U(1)$. Now $U(1) \times U(1) \subset SU(3)$ so once we modded out by the $U(1) \times U(1)$ we will 
be left with a single $U(1)$ that is in general a linear combination of the three $U(1)$s in $G$ which we parameterise by 
three integers $p$,$q$ and $r$ \footnote{The case where $p=q=0$ is the trivial fibration case where the coset becomes 
$\left[ SU(3)/U(1) \times U(1) \right]\times U(1)$. In that case this is the same as compactifying type IIA supergravity on the manifold
$SU(3)/U(1) \times U(1)$.}. We can repeat the analysis in the previous section and we find 
\ba
\label{cosetgenform}
g &=& a (e^1 \otimes e^1 +  e^2 \otimes e^2 ) + b (e^3 \otimes e^3 +  e^4
\otimes e^4 ) + c (e^5 \otimes e^5 +  e^6 \otimes e^6 ) + d e^7 \otimes e^7 \;
, \nonumber \\ 
\Psi_1 &=& \cs_1 e^7 \; , \nn \\
\Psi_2 &=& \cs_2 e^{12} + \cs_3 e^{34} + \cs_4 e^{56} \; , \\
\Psi_3 &=& \cs_5 \left( e^{135} + e^{146} - e^{236} + e^{245} \right) + 
\cs_6 \left( e^{136} - e^{145} + e^{235} + e^{246} \right) \; , \nn
\ea
Imposing the $SU(3)$ relations we arrive at equation (\ref{coset2su3}) where the basis forms explicitely read
\ba
z &\equiv& e^7 \; , \nn \\
\omega_1 &\equiv& -e^{12} \;\;\; \omega_2 \equiv e^{34} \;\;\; \omega_3 = -
e^{56} \; , \\ 
\tilde{\omega}^1 &\equiv& -e^{3456} \;\;\; \tilde{\omega}^2 \equiv e^{1256}
\;\;\; \tilde{\omega}^3 = - e^{1234} \; , \nn \\ 
\alpha_0 &\equiv& \left( -e^{136} +e^{145} -e^{235} -e^{246} \right) \;\;\;
\beta^0 \equiv -\quarter \left( e^{135} + e^{146} - e^{236} + e^{245}\right)
\; , \nn
\ea 
The differential relations on these basis forms are derived from 
\ba
de^1 &=& \alpha e^{72} - \half e^{36} + \half e^{45}\; , \nn \\
de^2 &=& \alpha e^{17} - \half e^{35} - \half e^{46} \; , \nn \\ 
de^3 &=& \beta e^{74} + \half e^{25} + \half e^{16} \; , \nn \\ 
de^4 &=& \beta e^{37} - \half e^{15} + \half e^{26}  \; , \\ 
de^5 &=& -\gamma e^{67} + \half e^{14} - \half e^{24}  \; , \nn\\
de^6 &=& \gamma e^{57} - \half e^{13} - \half e^{24}  \; , \nn\\ 
de^7 &=& -\alpha e^{12} - \beta e^{34} - \gamma e^{56} \; . \nn
\ea
These then give the differential relations (\ref{coset2diffrel}) where we have defined the structure constants
\ba
\alpha &\equiv& f^7_{\;\;12} = \frac{q}{\sqrt{3p^2 + q^2}} \; , \nonumber \\
\beta &\equiv& f^7_{\;\;34} = \frac{3p+q}{2\sqrt{3p^2 + q^2}} \; , \\
\gamma &\equiv& f^7_{\;\;56} = \frac{3p-q}{2\sqrt{3p^2 + q^2}} \; . \nn
\ea


\begin{thebibliography}{}
\bibitem{Campbell:1984zc}
I.~C.~G.~Campbell and P.~C.~West,
``N=2 D = 10 Nonchiral Supergravity And Its Spontaneous Compactification,''
Nucl.\ Phys.\ B {\bf 243}, 112 (1984).

\bibitem{Huq:1983im}
M.~Huq and M.~A.~Namazie,
``Kaluza-Klein Supergravity In Ten-Dimensions,''
Class.\ Quant.\ Grav.\  {\bf 2}, 293 (1985)
[Erratum-ibid.\  {\bf 2}, 597 (1985)].

\bibitem{Giani:1984wc}
F.~Giani and M.~Pernici,
``N=2 Supergravity In Ten-Dimensions,''
Phys.\ Rev.\ D {\bf 30}, 325 (1984).

\bibitem{Lukas:1998yy}
A.~Lukas, B.~A.~Ovrut, K.~S.~Stelle and D.~Waldram,
``The universe as a domain wall,''
Phys.\ Rev.\ D {\bf 59} (1999) 086001
[arXiv:hep-th/9803235].

\bibitem{Papadopoulos:1995da}
G.~Papadopoulos and P.~K.~Townsend,
``Compactification of D = 11 supergravity on spaces of exceptional holonomy,''
Phys.\ Lett.\ B {\bf 357}, 300 (1995)
[arXiv:hep-th/9506150].

\bibitem{Grana:2005jc}
  M.~Grana,
  ``Flux compactifications in string theory: A comprehensive review,''
  arXiv:hep-th/0509003.

\bibitem{Acharya:2002kv}
  B.~S.~Acharya,
  ``A moduli fixing mechanism in M theory,''
  arXiv:hep-th/0212294.

\bibitem{deCarlos:2004ci}
  B.~de Carlos, A.~Lukas and S.~Morris,
  ``Non-perturbative vacua for M-theory on G(2) manifolds,''
  JHEP {\bf 0412} (2004) 018
  [arXiv:hep-th/0409255].

\bibitem{Lambert:2005sh}
 N.~Lambert,
 ``Flux and Freund-Rubin superpotentials in M-theory,''
 arXiv:hep-th/0502200.

\bibitem{Gauntlett:2002sc}
 J.~P.~Gauntlett, D.~Martelli, S.~Pakis and D.~Waldram,
 ``G structures and wrapped NS5-branes,''
 Commun.\ Math.\ Phys.\  {\bf 247}, 421 (2004)
 [arXiv:hep-th/0205050].

\bibitem{Gauntlett:2003cy}
  J.~P.~Gauntlett, D.~Martelli and D.~Waldram,
  ``Superstrings with intrinsic torsion,''
  Phys.\ Rev.\ D {\bf 69} (2004) 086002
  [arXiv:hep-th/0302158].

\bibitem{House:2004pm}
  T.~House and A.~Micu,
  ``M-theory compactifications on manifolds with G(2) structure,''
  arXiv:hep-th/0412006.

\bibitem{Dall'Agata:2005fm}
  G.~Dall'Agata and N.~Prezas,
  ``Scherk-Schwarz reduction of M-theory on G2-manifolds with fluxes,''
  JHEP {\bf 0510} (2005) 103
  [arXiv:hep-th/0509052].

\bibitem{D'Auria:2005rv}
  R.~D'Auria, S.~Ferrara and M.~Trigiante,
  ``Supersymmetric completion of M-theory 4D-gauge algebra from twisted tori and fluxes,''
  arXiv:hep-th/0511158.

\bibitem{Kaste:2003zd}
  P.~Kaste, R.~Minasian and A.~Tomasiello,
  ``Supersymmetric M-theory compactifications with fluxes on  seven-manifolds and G structures,''
  JHEP {\bf 0307} (2003) 004
  [arXiv:hep-th/0303127].

\bibitem{Dall'Agata:2003ir}
 G.~Dall'Agata and N.~Prezas,
 ``N = 1 geometries for M-theory and type IIA strings with fluxes,''
 Phys.\ Rev.\ D {\bf 69} (2004) 066004
 [arXiv:hep-th/0311146].

\bibitem{Behrndt:2004mx}
 K.~Behrndt and C.~Jeschek,
 ``Superpotentials from flux compactifications of M-theory,''
 Class.\ Quant.\ Grav.\  {\bf 21}, S1533 (2004)
 [arXiv:hep-th/0401019].

\bibitem{Lukas:2004ip}
 A.~Lukas and P.~M.~Saffin,
 ``M-theory compactification, fluxes and AdS(4),''
 Phys.\ Rev.\ D {\bf 71}, 046005 (2005)
 [arXiv:hep-th/0403235].

\bibitem{Behrndt:2004bh}
 K.~Behrndt and C.~Jeschek,
 ``Fluxes in M-theory on 7-manifolds: G2, SU(3) and SU(2) structures,''
 arXiv:hep-th/0406138.

\bibitem{Gauntlett:2004hs}
 J.~P.~Gauntlett, D.~Martelli, J.~Sparks and D.~Waldram,
 ``Supersymmetric AdS backgrounds in string and M-theory,''
 arXiv:hep-th/0411194.

\bibitem{Franzen:2005ve}
  A.~Franzen, P.~Kaura, A.~Misra and R.~Ray,
  ``Uplifting the Iwasawa,''
  arXiv:hep-th/0506224.

\bibitem{Behrndt:2005im}
  K.~Behrndt, M.~Cvetic and T.~Liu,
  ``Classification of supersymmetric flux vacua in M theory,''
  arXiv:hep-th/0512032.

\bibitem{House:2005yc}
  T.~House and E.~Palti,
  ``Effective action of (massive) IIA on manifolds with SU(3) structure,''
  Phys.\ Rev.\ D {\bf 72} (2005) 026004
  [arXiv:hep-th/0505177].

\bibitem{Behrndt:2004km}
  K.~Behrndt and M.~Cvetic,
  ``General N=1 supersymmetric flux vacua of (massive) type IIA string theory''
  Phys.\ Rev.\ Lett.\ {\bf 95} (2005) 021601
  [arXiv:hep-th/0403049].

\bibitem{Behrndt:2004mj}
  K.~Behrndt and M.~Cvetic,
  ``General N=1 supersymmetric fluxes in massive type IIA string theory''
  Nucl.\ Phys.\ B {\bf 708} (2005) 45
  [arXiv:hep-th/0407263].

 \bibitem{Lust:2004ig}
  D.~Lust and D.~Tsimpis,
  ``Supersymmetric AdS(4) compactifications of IIA supergravity,''
  JHEP {\bf 0502} (2005) 027
  [arXiv:hep-th/0412250].

\bibitem{AFT}
R.~D'Auria, S.~Ferrara and M.~Trigiante,
``Spontaneously broken supergravity: Old and new facts,''
arXiv:hep-th/0512248.

\bibitem{joyce}
D.~Joyce
``Compactification of D = 11 supergravity on spaces of exceptional holonomy,''
Compact manifolds with special holonomy, OUP (2003)

\bibitem{FR}
P.~G.~O.~Freund and M.~A.~Rubin,
``Dynamics Of Dimensional Reduction,''
Phys.\ Lett.\ B {\bf 97} (1980) 233.

\bibitem{BJ}
K.~Behrndt and C.~Jeschek,
``Fluxes in M-theory on 7-manifolds: G structures and superpotential,''
Nucl.\ Phys.\ B {\bf 694} (2004) 99
[arXiv:hep-th/0311119].

\bibitem{Dall'Agata:2006vd}
  G.~Dall'Agata and S.~Ferrara,
  ``Updates in local supersymmmetry and its spontaneous breaking,''
  arXiv:hep-th/0601138.

\bibitem{FKMS}
T.~Friedrich, I.~Kath, A.~Moroianu and U.~Semmelmann,
``On nearly parallel G(2) structures,'' SFB-288-162, J.~Geom.~Phys.~{\bf 23}
(1997), 259.

\bibitem{JP}
J.~Polchinski, 
``String Theory'', 2 vols.~ Cambridge University Press (1998).

\bibitem{deWit:1984pk}
 B.~de Wit and A.~Van Proeyen,
 ``Potentials And Symmetries Of General Gauged N=2 Supergravity - Yang-Mills
 Models,''
 Nucl.\ Phys.\ B {\bf 245} (1984) 89.

\bibitem{Andrianopoli:1996vr}
 L.~Andrianopoli, M.~Bertolini, A.~Ceresole, R.~D'Auria, S.~Ferrara and P.~Fr\'{e},
  ``General Matter Coupled N=2 Supergravity,''
  Nucl.\ Phys.\ B {\bf 476}, 397 (1996)
  [arXiv:hep-th/9603004].
  
\bibitem{Andrianopoli:1996cm}
L.~Andrianopoli, M.~Bertolini, A.~Ceresole, R.~D'Auria, S.~Ferrara, P.~Fre and T.~Magri,
``N = 2 supergravity and N = 2 super Yang-Mills theory on general scalar
 manifolds: Symplectic covariance, gaugings and the momentum map,''
J.\ Geom.\ Phys.\  {\bf 23} (1997) 111
[arXiv:hep-th/9605032].

\bibitem{DAASV}
G.~Dall'Agata, R.~D'Auria, L.~Sommovigo and S.~Vaula,
``D = 4, N = 2 gauged supergravity in the presence of tensor multiplets,''
Nucl.\ Phys.\ B {\bf 682} (2004) 243
[arXiv:hep-th/0312210].

\bibitem{SV}
L.~Sommovigo and S.~Vaula,
``D = 4, N = 2 supergravity with Abelian electric and magnetic charge,''
Phys.\ Lett.\ B {\bf 602} (2004) 130
[arXiv:hep-th/0407205].

\bibitem{ASV}
R.~D'Auria, L.~Sommovigo and S.~Vaula,
``N = 2 supergravity Lagrangian coupled to tensor multiplets with electric and
magnetic fluxes,''
JHEP {\bf 0411} (2004) 028
[arXiv:hep-th/0409097].

\bibitem{Grana:2005ny}
  M.~Grana, J.~Louis and D.~Waldram,
  ``Hitchin functionals in N = 2 supergravity,''
  arXiv:hep-th/0505264.

\bibitem{Hitchin}
N.~J.~Hitchin,
``Stable forms and special metrics,''
arXiv:math.dg/0107101. \\
N.~J.~Hitchin,
``Generalized Calabi-Yau manifolds,''
Quart.\ J.\ Math.\ Oxford Ser.\  {\bf 54} (2003) 281
[arXiv:math.dg/0209099].

\bibitem{Gurrieri:2002wz}
  S.~Gurrieri, J.~Louis, A.~Micu and D.~Waldram,
  ``Mirror symmetry in generalized Calabi-Yau compactifications,''
  Nucl.\ Phys.\ B {\bf 654} (2003) 61
  [arXiv:hep-th/0211102].

\bibitem{Tomasiello:2005bp}
  A.~Tomasiello,
  ``Topological mirror symmetry with fluxes,''
  arXiv:hep-th/0502148.

\bibitem{D'Auria:2004tr}
 R.~D'Auria, S.~Ferrara, M.~Trigiante and S.~Vaula,
 ``Gauging the Heisenberg algebra of special quaternionic manifolds,''
 Phys.\ Lett.\ B {\bf 610}, 147 (2005)
 [arXiv:hep-th/0410290].

\bibitem{BW}
C.~Beasley and E.~Witten,
``A note on fluxes and superpotentials in M-theory compactifications on
manifolds of G(2) holonomy,''
JHEP {\bf 0207} (2002) 046
[arXiv:hep-th/0203061].

\bibitem{KS}
J.~L.~P.~Karthauser and P.~M.~Saffin,
``The dynamics of coset dimensional reduction,''
arXiv:hep-th/0601230.

\bibitem{Cecotti:1984rk}
 S.~Cecotti, L.~Girardello and M.~Porrati,
 ``Two Into One Won't Go,''
 Phys.\ Lett.\ B {\bf 145}, 61 (1984).

\bibitem{Cecotti:1985sf}
 S.~Cecotti, L.~Girardello and M.~Porrati,
 ``An Exceptional N=2 Supergravity With Flat Potential And Partial Superhiggs,''
 Phys.\ Lett.\ B {\bf 168}, 83 (1986).

\bibitem{Ferrara:1995xi}
 S.~Ferrara, L.~Girardello and M.~Porrati,
 ``Spontaneous Breaking of N=2 to N=1 in Rigid and Local Supersymmetric
 Theories,''
 Phys.\ Lett.\ B {\bf 376} (1996) 275
 [arXiv:hep-th/9512180].

\bibitem{Andrianopoli:2001gm}
  L.~Andrianopoli, R.~D'Auria and S.~Ferrara,
  ``Consistent reduction of N = 2 $\to$ N = 1 four dimensional supergravity coupled to matter,''
  Nucl.\ Phys.\ B {\bf 628}, 387 (2002)
  [arXiv:hep-th/0112192].

\bibitem{Louis:2002vy}
  J.~Louis,
  ``Aspects of spontaneous N = 2 $\to$ N = 1 breaking in supergravity,''
  arXiv:hep-th/0203138.

\bibitem{Gunara:2003td}
 B.~E.~Gunara,
 ``Spontaneous N=2 $\to$ N=1 supersymmetry breaking and the super-Higgs effect
 in supergravity,''
 href{http://www.slac.stanford.edu/spires/find/hep/www?irn=6000282}
 {\ (SPIRES entry)} and references therein.

\bibitem{Micu:2004tz}
  A.~Micu,
  ``Heterotic compactifications and nearly-Kaehler manifolds,''
  Phys.\ Rev.\ D {\bf 70} (2004) 126002
  [arXiv:hep-th/0409008].

\bibitem{Castellani:1983tc}
L.~Castellani and L.~J.~Romans,
``N=3 And N=1 Supersymmetry In A New Class Of Solutions For D = 11
Supergravity,''
Nucl.\ Phys.\ B {\bf 238} (1984) 683.

\bibitem{ABL}
L.~B.~Anderson, A.~B.~Barrett and A.~Lukas,
``M-Theory on the Orbifold C**2/Z(N),''
arXiv:hep-th/0602055.

  \bibitem{Mueller-Hoissen:1987cq}
  F.~Mueller-Hoissen and R.~Stuckl,
  ``Coset Spaces And Ten-Dimensional Unified Theories,''
  Class.\ Quant.\ Grav.\  {\bf 5} (1988) 27.

\bibitem{AZ}
  L.~Anguelova and K.~Zoubos,
  ``Flux superpotential in heterotic M-theory,''
  arXiv:hep-th/0602039.

\bibitem{CL}
  M.~Cvetic and T.~Liu,
  ``Moduli Stabilization in M-theory with SU(3) structures,''
  private communication.

\end{thebibliography}
\end{document}